\documentclass[reprint,amsmath,amssymb,aps]{revtex4-1}

\usepackage{graphicx}
\usepackage{bm}
\usepackage{hyperref}
\usepackage{amsmath}
\usepackage{braket}
\usepackage{dsfont}		%for symbols of number systems
\usepackage{siunitx}	%SI units

\newcommand{\ee}{\mathrm{e}}
\newcommand{\ii}{\mathrm{i}}

\begin{document}

	\title{Helicity flip of high-harmonic photons in Haldane nanoribbons}

	\author{Christoph J\"ur\ss}
	\affiliation{Institute of Physics, University of Rostock, 18051 Rostock, Germany }
	\author{Dieter Bauer}
	\affiliation{Institute of Physics, University of Rostock, 18051 Rostock, Germany }
	\date{\today}
	
	\begin{abstract}
          Recent studies in high-harmonic spectroscopy of condensed matter mainly focused on the bulk of the system under consideration. In this work, we investigate the response of thin, hexagonal nanoribbons to an intense laser pulse that is linearly polarized along the ribbon. Such nanoribbons are prime examples of two-dimensional systems that are bulk-like in one direction and finite in the other direction. Despite the atomically thin scale in the direction perpendicular to the linearly polarized driving laser field, the emitted harmonics are elliptically polarized if an alternating onsite potential and Haldane hopping is taken into account. For given hoppings,  we find a sudden change of the helicity for a certain harmonic order. The origin of this flip is traced back to phase differences between the components of Bloch states.  
	\end{abstract}

	\maketitle

	\section{Introduction} 
	
	High-harmonic generation (HHG) in condensed matter is a relatively new though meanwhile intensely investigated  topic in strong-field,  attosecond physics, as it allows for an all optical probing of solids \cite{Ghimire2011,SchubertO.2014,VampaPhysRevLett.115.193603,Hohenleutner2015,Luu2015,ndabashimiye_solid-state_2016,LangerF.2017,TancPhysRevLett.118.087403,you_high-harmonic_2017,Zhang2018,Vampa2018,Baudisch2018,Garg2018,Abadie2018,Yue2020}. 
	Topological insulators \cite{topinsRevModPhys.82.3045,topins,topinsshortcourse} are an especially  interesting kind of condensed matter because they can host edge currents that are immune to scattering thanks to their ``topological protection.'' The steering of these ballistic edge currents by light may pave the way towards ultrafast electronics \cite{Reimann2018}. Recently, the study of HHG in topological insulators started both theoretically \cite{PhysRevB.96.075409,bauer_high-harmonic_2018,Silva2019,chacon_observing_2018,DrueekeBauer2019,JuerssBauer2019} and experimentally \cite{Luu2018,Reimann2018}. 
	
	In two dimensions, graphene is one of the most investigated condensed matter systems. Haldane introduced a toy model \cite{Haldane_1988} to make graphene topological by adding (i) an alternating onsite potential to open a band gap, and (ii) a complex next-nearest neighbor hopping (which has an effect similar to a magnetic field). The system was implemented experimentally using, e.g., cold atoms \cite{Jotzu2014}. HHG  in the bulk of ``Haldanite''  coupled to a laser  field was studied recently. It was found that the topological phase of the system determines the helicity of the emitted photons \cite{Silva2019}, and that the topological phase can be measured through   circular dichroism \cite{chacon_observing_2018}. Besides graphene bulk, the electronic structure and topological properties of graphene nanoribbons were studied as well, see, e.g., \cite{Cao2017,PANTALEON2018191}. The ribbons are finite and hence edge effects become important. 

		In this work we investigate ``zig-zag'' graphene-like (i.e., with Haldane hopping) nanoribbons in laser fields linearly polarized along the ribbon. The structure and notation is introduced in Fig. \ref{fig1}.
	The electrons can hop in two spatial dimensions $x$ and $y$ but as the ribbons are much longer (in $x$) than wide (in $y$) the system is almost one-dimensional. One unit cell $n$ contains four sublattice sites $\alpha = 1,2,3,4$, as shown in Fig. \ref{fig1} (a). Finite sized ribbons with periodic boundary conditions in $x$ direction are investigated in tight-binding approximation. Hence the right most unit cell $n=N$ is connected with the left most one $n=1$.  In addition to the usual real-valued hopping amplitude between neighboring atoms, an additional, alternating onsite potential and a complex next-nearest neighbor hopping as in the Haldane model is included. The system is topologically non-trivial for a sufficiently large next-nearest neighbor hopping. 
	 
	In the following Section \ref{sec:theory}, the tight-binding modelling is introduced, including the coupling to an external field in  \ref{sec:coupling} and the properties of the bulk system (with respect to the $x$ direction) in \ref{sec:bulk_theory}. The results for the bulk system  are used to explain the features observed in the HHG spectra in Section \ref{sec:HHG}. The results are compared to the respective system without periodic boundary conditions in Section  \ref{sec:ribbon_edges}.  We conclude in Section \ref{sec:summandout}. Details on the calculation of the current and its dependence on phase-differences between Bloch-state components are given in Appendix \ref{app:A}. Atomic units ($\hbar=|e|=m_e=4\pi\epsilon_0=1$) are used throughout the paper if not stated otherwise.
	
	\section{Theory}\label{sec:theory}

		\subsection{System without external field}
		
		\begin{figure} 
			\includegraphics[width = \columnwidth]{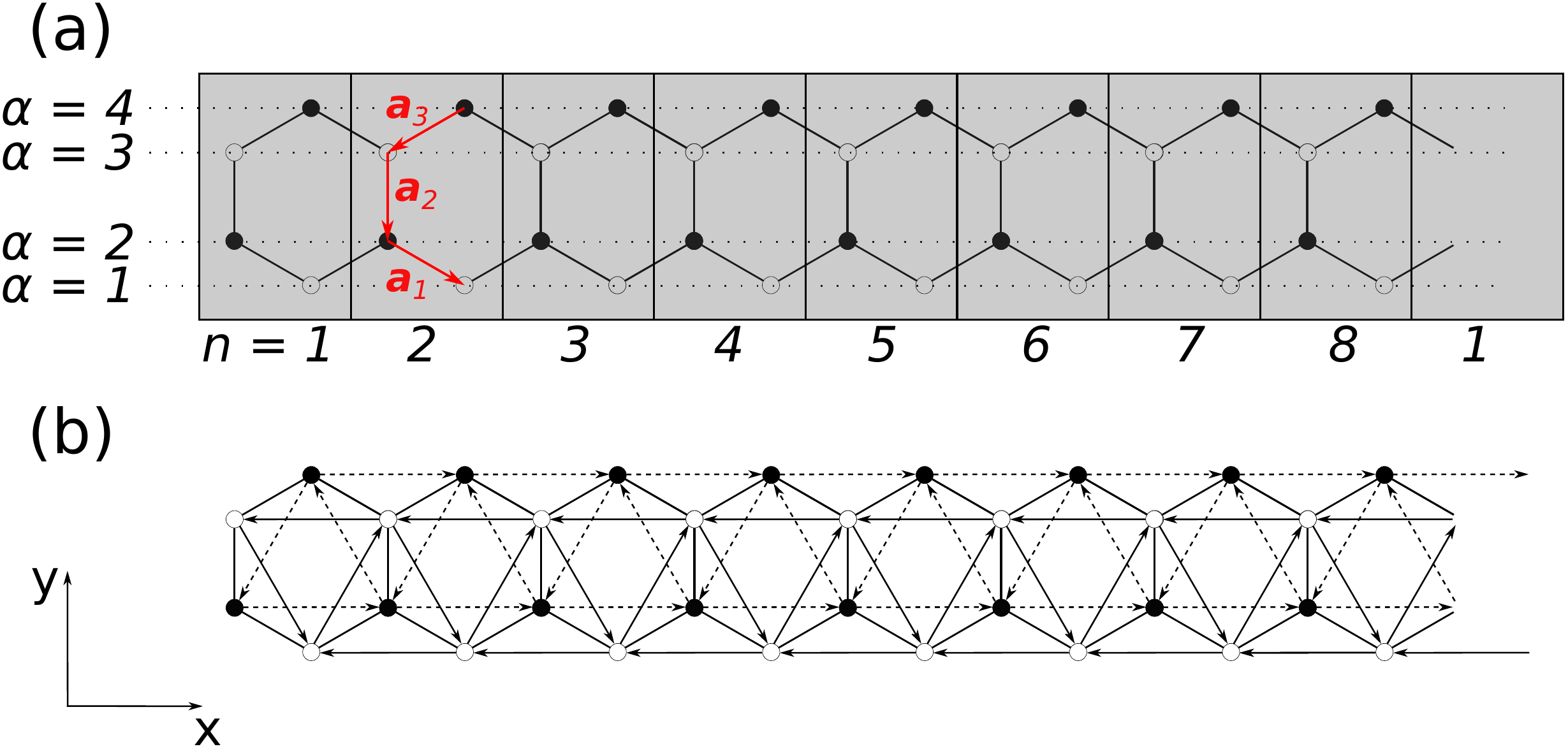}
			\caption{\label{fig1} (a) Sketch of the hexagonal ribbon in zig-zag configuration. Sites with on-site potential $M$ ($-M$) are indicated by open (filled) circles. Solid lines show nearest neighbor hoppings with amplitude $t_1\in\mathds{R}$ between adjacent sites. 
			The unit cells are numbered by $n=1,2,...,N$ (here $N = 8$). The unit cell $N$ is connected with unit cell $n = 1$, implying periodic boundary conditions. The sub-lattice sites are denoted by $\alpha$. (b) Hexagonal ribbon with an additional next-nearest neighbor hopping of strength $t_2\mathrm{e}^{\mathrm{i}\phi}$ along the arrows and $t_2\mathrm{e}^{-\mathrm{i}\phi}$ in the opposite direction ($t_2 \in \mathds{R}$). }
		\end{figure}
		
		Hexagonal ribbons with zig-zag edges as sketched in Fig. \ref{fig1} are investigated. The circles in Fig. \ref{fig1}(a) indicate the atomic positions, and the vectors $\bm{a}_i$ ($i=1,2,3$), 
		\begin{align}
		\bm{a}_1 = a/2\left(\sqrt{3}, -1\right)^\top, \nonumber\\
		\bm{a}_2 = a\left(0, -1\right)^\top,\\
		\bm{a}_3 = a/2\left(-\sqrt{3}, -1\right)^\top,\nonumber
		\end{align}
		connect neighboring sites.
		A tight binding approach is used. The hopping amplitude between nearest neighbors is given by $t_1$ and sketched by solid lines in Fig. \ref{fig1} (a,b). With only nearest-neighbor hopping, this system describes a graphene ribbon. Bulk graphene is a conductor and has a vanishing band gap. If an alternating on-site potential ($\pm M$) is introduced, a band gap between valence and conduction band opens, and the system becomes an insulator. This can be realized by using two different elements instead of carbon only in the case of graphene, for instance boron and nitrogen  in  hexagonal boron nitride (h-BN). The sites with an on-site potential of $M$ ($-M$) are denoted as sublattice sites A (B) and are indicated by  open (filled) circles in Fig. \ref{fig1}(a,b).
		
	Haldane proposed a way to make such a system topologically non-trivial \cite{Haldane_1988}. He introduced a complex hopping $t_2\mathrm{e}^{\mathrm{i}\phi}$ between next-nearest neighbors. Here, $t_2\in\mathds{R}$ is the hopping amplitude, and $\phi$ is the phase of the hopping. This hopping is sketched in Fig \ref{fig1}(b) with arrows. The orientation there denotes a hopping of $t_2\mathrm{e}^{\mathrm{i}\phi}$, the hopping in the opposite direction is  $t_2\mathrm{e}^{-\mathrm{i}\phi}$. 
	The unit cells are numbered by the index $n = 1,2,3,...,N$. Each cell contains four atomic sites, labeled by $\alpha = 1,2,3,4$. Periodic boundary conditions are used. Hence to the right of hexagon $n=N$ follows the first hexagon $n=1$ again.

		An electronic  wavefunctions in tight-binding description has the form 
		\begin{align}
		\ket{\Psi} &= \sum_{n = 1}^{N}\sum_{\alpha = 1}^4 g_{n,\alpha}\ket{n,\alpha},
		\end{align}
		where  $\ket{N+1,\alpha} = \ket{1,\alpha}$.
		The Hamiltonian reads 
		\begin{align}
		\hat{H}_0 = \hat{H}_\mathrm{nn} + \hat{H}_M + \hat{H}_\mathrm{nnn}
		\end{align} 
		where $\hat{H}_\mathrm{nn}$, $\hat{H}_M$ and $\hat{H}_\mathrm{nnn}$ are the Hamiltonians describing the nearest neighbor hopping, the on-site potential and the next-nearest neighbor hopping, respectively.
		The nearest neighbor part reads	
		\begin{align}\label{eq:Hamiltonian_nn_periodic}
		\hat{H}_\mathrm{nn} &= t_1\sum_{n = 1}^N\left(\left[\sum_{\alpha = 1}^3\ket{n,\alpha}\bra{n,\alpha + 1}\right] + \ket{n+1,3}\bra{n,4}\right. \nonumber\\
		& + \ket{n+1,2}\bra{n,1}\bigg) +\mathrm{h.c.},
		\end{align}
		the onsite part reads
		\begin{align}
		\hat{H}_M^{\mathrm{periodic}} &= M\sum_{n = 1}^N\sum_{\alpha = 1}^4 \left(-1\right)^{\alpha + 1}\ket{n,\alpha}\bra{n,\alpha},
		\end{align}
		and the next-nearest neighbor Hamiltonian is
		\begin{align}
		\hat{H}_\mathrm{nnn} &= t_2 \sum_{n = 1}^N\bigg(\mathrm{e}^{\mathrm{i}\phi}\Big[\ket{n,2}\bra{n,4} + \ket{n,1}\bra{n,3} \nonumber\\
		& +\ket{n+1,3}\bra{n,1}+\ket{n,4}\bra{n+1,2} \Big] \nonumber\\
		& + \sum_{\alpha = 1}^4\mathrm{exp}\Big\{\left(-1\right)^\alpha \mathrm{i}\phi\Big\}\ket{n+1,\alpha}\bra{n,\alpha}\bigg) +\mathrm{h.c.}.
		\end{align}

		The time-independent Schr\"odinger equation (TISE) 
		\begin{align}\label{eq:TISE}
			\hat{H}_0 \ket{\psi_i}= E_i \ket{\psi_i} 
		\end{align}
		is solved to obtain the eigenstates of the system. Here, $E_i$ is the eigenvalue of state $\ket{\psi_i}$. The ribbon contains $L = 4N $ sites, hence the Hamiltonian has $L$ orthogonal eigenstates.

		\subsection{Coupling to an external field}\label{sec:coupling}
		
		For the coupling to an external field, the dipole approximation and velocity gauge are used. The elements of the Hamiltonian matrix become time-dependent according \cite{Graf_1995}
		\begin{align}
			\bra{n,\alpha}\hat{H}(t)\ket{n',\alpha'} = \bra{n,\alpha}\hat{H}_0\ket{n',\alpha'}\mathrm{e}^{-\mathrm{i}\left(\bm{r}_{n,\alpha} - \bm{r}_{n',\alpha'}\right)\cdot\bm{A}(t)},
		\end{align} 
		where  $\bm{r}_{n,\alpha}$ ($\bm{r}_{n',\alpha'}$) are the positions of the atoms at hexagon $n$ ($n'$) and site $\alpha$ ($\alpha'$), and $\bm{A}(t)$ is the vector potential of the $n_{cyc}$-cycle laser pulse of frequency $\omega_0$ and amplitude $A_0$,
		\begin{align}
		\bm{A}(t) = \left(A(t),0\right)^\top~\mathrm{and}~A(t)= A_0~\sin ^2\left(\frac{\omega_0 t}{2 n_{cyc}}\right)~\sin (\omega_0 t)
		\end{align}
		for $0\leq t \leq 2\pi n_{cyc}/\omega_0$ and zero otherwise. The parameters used in the following are $A_0 = 0.05$ (corresponding to an intensity of $\simeq 5\times 10^{9}~ \mathrm{Wcm}^{-2}$),  $\omega_0 = 7.5\cdot 10^{-3}$ (i.e., $\lambda = \SI{6.1}{\micro\meter}$), and  $n_{cyc} = 5$.

		The system contains multiple electrons but we neglect electron-electron interaction. A single-electron wavefunction $\ket{\Psi^i(t)}$  can be expanded in the eigenstates from the TISE (\ref{eq:TISE}),
		\begin{equation}\label{eq:expansion_psi}
		\ket{\Psi^i(t)} = \sum_{l=0}^{L-1} c_l^i(t)  \ket{\psi_l} e^{-\mathrm{i} E_l t}.
		\end{equation}
		The initial conditions are chosen
		\begin{align}
			\ket{\Psi^i(t=0)} = \ket{\psi_i}.
		\end{align}
		It is assumed that all eigenstates below the Fermi-level ($E_i < 0$) are occupied before the laser hits the nanoribbon, i.e.,  the lower half of the states are occupied initially ($i = 0,1,2,...,L/2-1$). 
		
		The time-dependent Hamiltonian can be written in the form 
		\begin{align}\label{eq:TDHam}
			\hat{H}(t) = \hat{H}_0 + \hat{V}(t).
		\end{align}
		The ansatz (\ref{eq:expansion_psi}) is plugged into the time-dependent Schr\"odinger equation (TDSE)		
		\begin{equation}\label{eq:TDSE}
		\mathrm{i}\partial_t \ket{\Psi(t)} = \hat{H}(t) \ket{\Psi(t)} = \left(\hat{H}_0 + \hat{V}(t) \right)\ket{\Psi(t)}	.
		\end{equation} 
		After a few steps, one arrives at a system of differential equations for the coefficients $c_j^i$,		
		\begin{equation}\label{eq:c_i}
		\dot{c}_j^i(t) = -\mathrm{i}\sum_{l=0}^{L-1} c_l^i(t)  e^{-\mathrm{i}(E_l -E_j) t}\bra{\psi_j}\hat{V}(t)  \ket{\psi_l}.
		\end{equation}

		The current is required to calculate the harmonic spectra. The current operator reads \cite{Review_Transport}
		\begin{align}\label{eq:current_operator}
			\hat{\bm{j}}(t) = -\mathrm{i}\sum_{n,\alpha}\sum_{n',\alpha'}\left(\bm{r}_{n,\alpha} - \bm{r}_{n',\alpha'} \right)\ket{n,\alpha}H_{n,\alpha}^{n',\alpha'}(t)\bra{n',\alpha'},
		\end{align}
		with $H_{n,\alpha}^{n',\alpha'}(t) = \bra{n,\alpha}\hat{H}(t)\ket{n',\alpha'}$.
		The expectation value of the current is calculated as
		\begin{align}
			\bm{J}(t) = \sum_{i}\bra{\Psi^i(t)}\hat{\bm{j}}(t)\ket{\Psi^i(t)}
		\end{align}
		where the sum runs over all propagated electrons in the states states $\ket{\Psi^i(t)}$ (here $i=0,1,2,...,L/2-1$). The current has two components $\bm{J}(t) = \left(J_\parallel(t),J_\perp(t)\right)$. The vector potential is linearly polarized along the chain $\bm{A}(t) = \left(A(t),0\right)^\top$. The component $J_\parallel$ ($J_\perp$) refers to the polarization component parallel (perpendicular) to the vector potential. 
		
		The HHG spectrum is calculated by Fourier-transforming the current \cite{Bandrauk2009,Baggesen_2011,bauer_computational_2017},
		\begin{align}
			P_{\parallel,\perp}(\omega) = |P_{\parallel,\perp}(\omega)|\mathrm{e}^{\mathrm{i}\Phi_{\parallel,\perp}(\omega)}=  \mathrm{FFT}\left[J_{\parallel,\perp}(t)\right] .
		\end{align}
		The phase difference 
		\begin{align}\label{eq:phase}
			\Delta\Phi = \Phi_\parallel - \Phi_\perp
		\end{align} 
		indicates the polarization (i.e., helicity) of the emitted photons. 

		\subsection{Nanoribbon bulk Hamiltonian}\label{sec:bulk_theory}
		
		The Bloch  ansatz
		\begin{align}\label{eq:Bloch_ansatz}
		\ket{\psi_i} &= \frac{1}{\sqrt{N}}\sum_{m=1}^N \ee^{\ii m k_i d}\ket{m} \otimes \left(u_1(k_i)\ee^{\ii k_i d/2}\ket{1} \right.\nonumber\\
		&\left. + u_2(k_i)\ket{2} + u_3(k_i)\ket{3} + u_4(k_i)\ee^{\ii k_i d/2}\ket{4}\right)
		\end{align}
		can be used to simplify the TISE (\ref{eq:TISE}). The phase factors $\ee^{\ii k_i d/2}$ are included to take the shifts inside one unit cell into account. The function  $u_\alpha(k_i)$ is the periodic part (at site $\alpha$) of a Bloch state $\ket{\psi_i}$, and $d = \sqrt{3}a$ is the lattice constant. Inserting this Bloch ansatz into the TISE (\ref{eq:TISE}) yields, after a few standard calculation steps,
		\begin{align}
		\hat{H}_\mathrm{bulk}(k_i) \bm{u}(k_i) = E_i\bm{u}(k_i)
		\end{align}
		where 
		\begin{align}\label{eq:Hamiltonian_bulk}
		\hat{H}_\mathrm{bulk}(k_i) = \begin{pmatrix}
		M_+(k_i) & T_1(k_i) & h_-(k_i)&0\\
		T_1(k_i)&M_-(k_i)&t_1&h_+(k_i)\\
		h_-(k_i) & t_1 & M_+(k_i)&T_1(k_i)\\
		0&h_+(k_i)&T_1(k_i)&M_-(k_i)
		\end{pmatrix}
		\end{align}
		with 
		\begin{align}
		\bm{u}(k_i) &= \left(u_1(k_i),u_2(k_i),u_3(k_i),u_4(k_i)\right)^\top\\
		T_1(k_i) &= 2 t_1 \cos (k_i d/2),\\
		M_\pm(k_i) &= \pm M + 2 t_2\cos \left(\phi \pm k_i d \right),\\
		h_\pm(k_i) &=  2 t_2\cos \left(\phi\pm k_i d/2\right)
		\end{align}
		is the bulk Bloch Hamiltonian for the nanoribbon. 
		
		The bulk Hamiltonian is a $4\times4$ matrix so that there are four solutions of the TISE for each $k_i$, i.e.,  four bands. The bands are indicated by $j = 1,2,3,4$ with $E_i^j$ and $\bm{u}^j(k_i) = \left(u_1^j(k_i),u_2^j(k_i),u_3^j(k_i),u_4^j(k_i)\right)^\top$. The states are sorted so that the energies are in  ascending order, $E_i^1 \leq E_i^2 \leq E_i^3 \leq E_i^4 $. The eigenvector components can be written as $u_\alpha^j(k_i) = \left|u_\alpha^j(k_i)\right|\ee^{\ii\phi_\alpha^j(k_i)}$. The phases $\phi_\alpha^j(k_i)$ are random because the states $u_\alpha^j(k_i)$ are calculated for each $k_i$ separately. In order to compare the phases, a certain structure gauge is applied, 
		\begin{align}
			\tilde{\bm{u}}^j(k_i) &= \bm{u}^j(k_i)\ee^{-\ii \phi_{1}^j(k_i)} ,\\
			\tilde{u}^j_\alpha(k_i) &= \left|u_\alpha^j(k_i)\right|\ee^{\ii\left(\phi_\alpha^j(k_i)-\phi_{1}^j(k_i)\right)}      = \left|u_\alpha^j(k_i)\right|\ee^{\ii\phi'_{\alpha,j}(k_i)} . 
		\end{align}
		By definition, the phases $\phi'_{1,j}(k_i)$ are zero in this gauge. 
		Observables such as HHG spectra must be independent of the gauge. And indeed, for the generation of high-harmonics, only phase differences between bands are important, 
		\begin{align}
			\Delta \phi^{j,j'}_{\alpha,\alpha'} = \phi'_{\alpha,j} - \phi'_{\alpha',j'}.
		\end{align}
		Note, that the Hamiltonian (\ref{eq:Hamiltonian_bulk}) is symmetric and real, which means that  the $\tilde{u}^j_\alpha(k_i)$ are also real, and the phases $\Delta \phi^{j,j'}_{\alpha,\alpha'}$ can only be $0$ or $\pi$ (we restrict the phases to $\Delta \phi^{j,j'}_{\alpha,\alpha'}\in\big[0,2\pi\big[$).
		As it will be demonstrated in Section \ref{sec:part_filled_bands}, states near the minimal band gap contribute most to the spectrum. Hence the phase difference between the bands around the band gap, $\Delta \phi'_{\alpha}\equiv\Delta \phi^{3,2}_{\alpha,\alpha}$, is investigated in detail.
		
	\section{Results for the periodic System}
		
	The system is initialized with the following parameters: the distance between adjacent atoms is $a = 2.68 \simeq  1.42 ~\mathrm{\AA}$, and the hopping between these atoms is set to $t_1 = -2.7 ~\mathrm{eV} \approx -0.1~(\mathrm{a.u.})$, which are the known parameters of graphene \cite{Cooper_2012}. In the bulk, a topological phase transition occurs at $t_2 = \pm M/(3\sqrt{3} ~\sin \phi)$ \cite{Haldane_1988}. This formula is not fulfilled exactly for theses ribbons. But still, the on-site potential is chosen to be rather small, $M = 0.01$, to assure a topological phase transition for a small next-nearest neighbor hopping. In particular, the transition should be observed for $|t_2| < |t_1|$ because hopping between next-nearest neighbors should be smaller than hopping between nearest ones. We set the next-nearest neighbor hopping to a purely imaginary number (i.e., $\phi = \pi/2$).
	The system contains $N=30$ atoms.

	\subsection{The bulk system}\label{sec:bulk}
	
\begin{figure}
			\centering
			\includegraphics[width=\columnwidth]{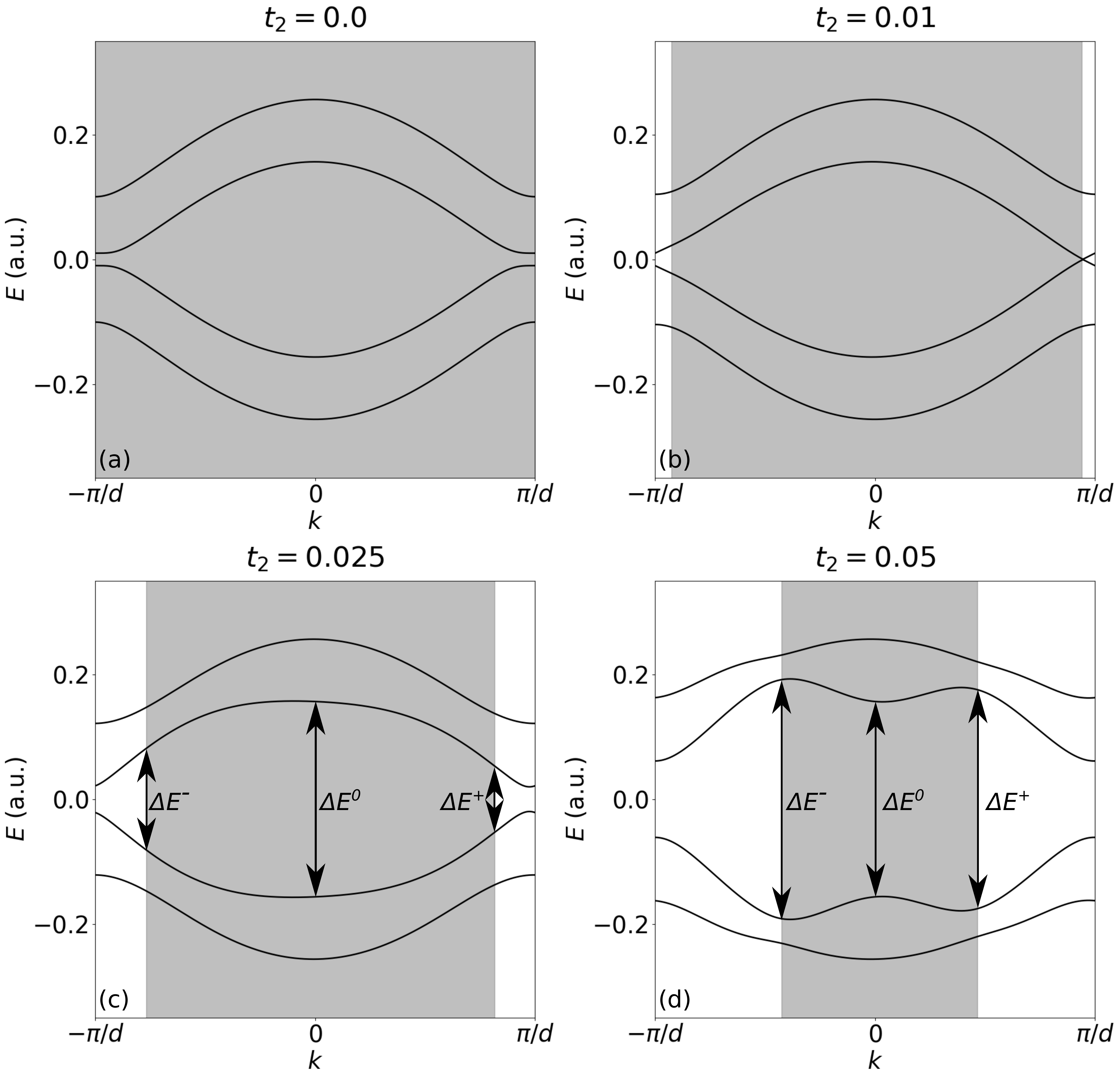}
			\caption{Band structure of the nanoribbon with periodic boundary in $x$ direction for different $t_2$. The shaded areas are identical to the ones in Fig. \ref{fig:phases_Bloch}.}
			\label{fig:Bands}
		\end{figure}

\begin{figure}
			\centering
			\includegraphics[width=\columnwidth]{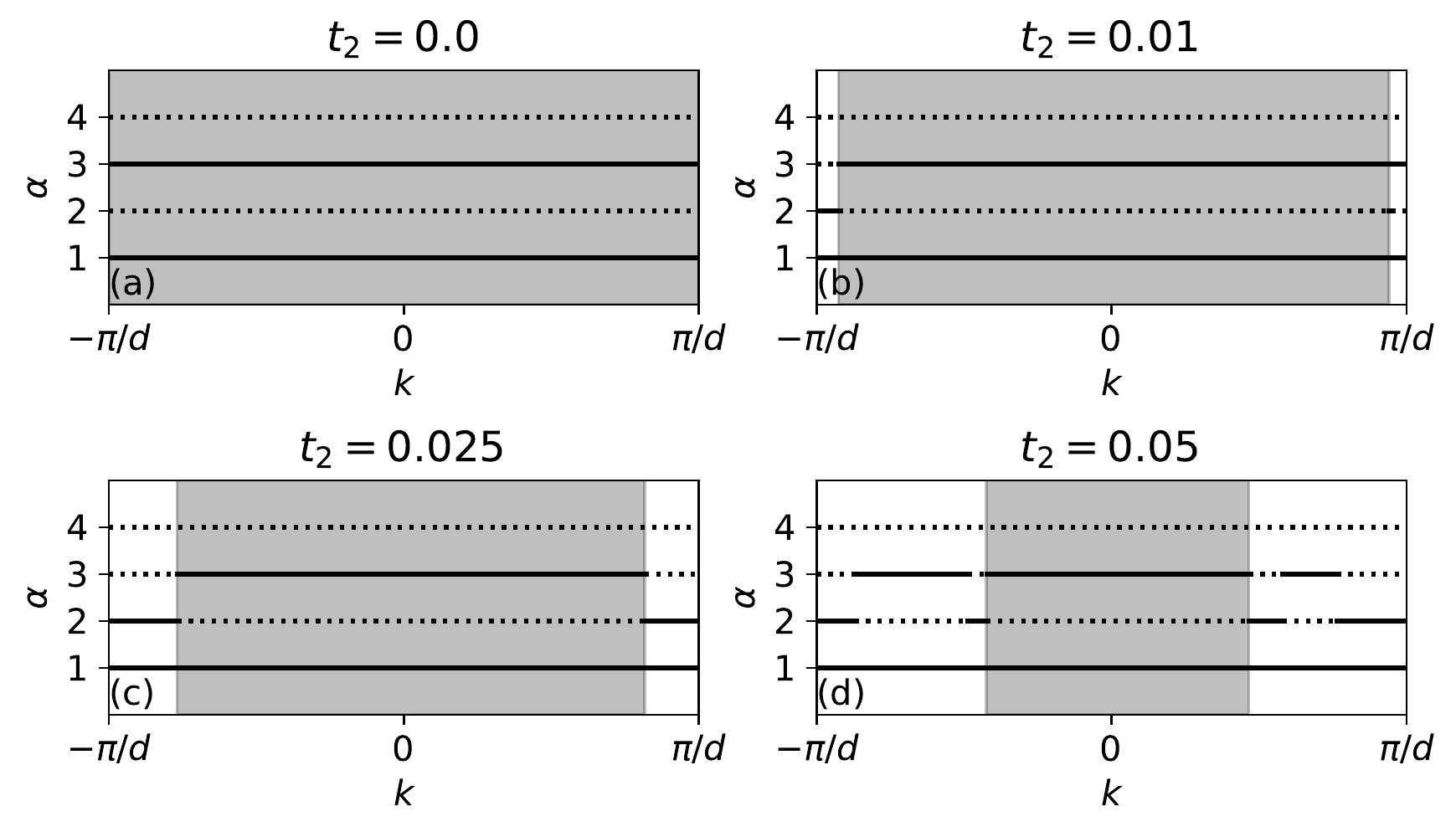}
			\caption{Plots of $\Delta \phi'_{\alpha}$ as function of $k$ for various $t_2$. This phase difference can assume two different values only. Solid lines denote  $\Delta \phi'_{\alpha} = 0$, dotted lines $\Delta \phi'_{\alpha} = \pi$. The gray shaded areas indicate the region around $k = 0$ where the phase differences are identical to the $t_2 = 0$-case.}
			\label{fig:phases_Bloch}
		\end{figure}

		In Fig.\ \ref{fig:Bands}, the bands are plotted for four different $t_2$. There are four bands that are all well separated at every $k$ for $t_2 = 0$. The minimal band gap between bands $j=2$ and $3$ is located at the boundaries of the first Brillouin zone ($k = \pm\pi/d$). 
		This band gap is $\Delta E_\mathrm{gap}(t_2 = 0) = 2|M| = 0.02$.
		For $t_2 \simeq 0.01$ the gap between bands $j=2$ and $3$ vanishes and the topological phase transition occurs. For larger $t_2$, the band gap opens again. At the two boundaries of the gray-shaded areas in Fig.\ \ref{fig:Bands} phase differences $\Delta \phi'_{\alpha}\equiv\Delta \phi^{3,2}_{\alpha,\alpha}$ of the periodic part of the Bloch functions jump (at least for one $\alpha$). The phase differences are plotted in Fig. \ref{fig:phases_Bloch}. Due to the real, symmetric Hamiltonian the phase differences can only be $\Delta \phi'_{\alpha} = 0$ (solid line) or $\Delta \phi'_{\alpha} = \pi$ (dotted line). 
		
		For vanishing next-nearest neighbor hopping $t_2 = 0$, the phase differences $\Delta\phi'_\alpha$ are constant over the whole Brillouin zone, see Fig. \ref{fig:phases_Bloch} (a). Their values are $0$, $\pi$, $0$ and $\pi$ for sites $1$, $2$, $3$ and $4$, respectively. As $t_2$ increases, at least one value starts to differ around the boundaries of the Brillouin zone ($k = \pm\pi/d$) while around $k=0$ they remain identical to the case $t_2 = 0$. The gray shaded areas in Fig. \ref{fig:phases_Bloch} indicate the regions of $t_2=0$-like behavior around $k = 0$. The gray shaded areas shrink towards $k=0$ with increasing $t_2$. Note that $\Delta\phi'_2$ jumps at $t_2 =0.01$ on the right edge of the shaded area from $\pi$ to $0$ and back to $\pi$ at a slightly larger $k$.

\begin{figure}
			\centering
			\includegraphics[width=\columnwidth]{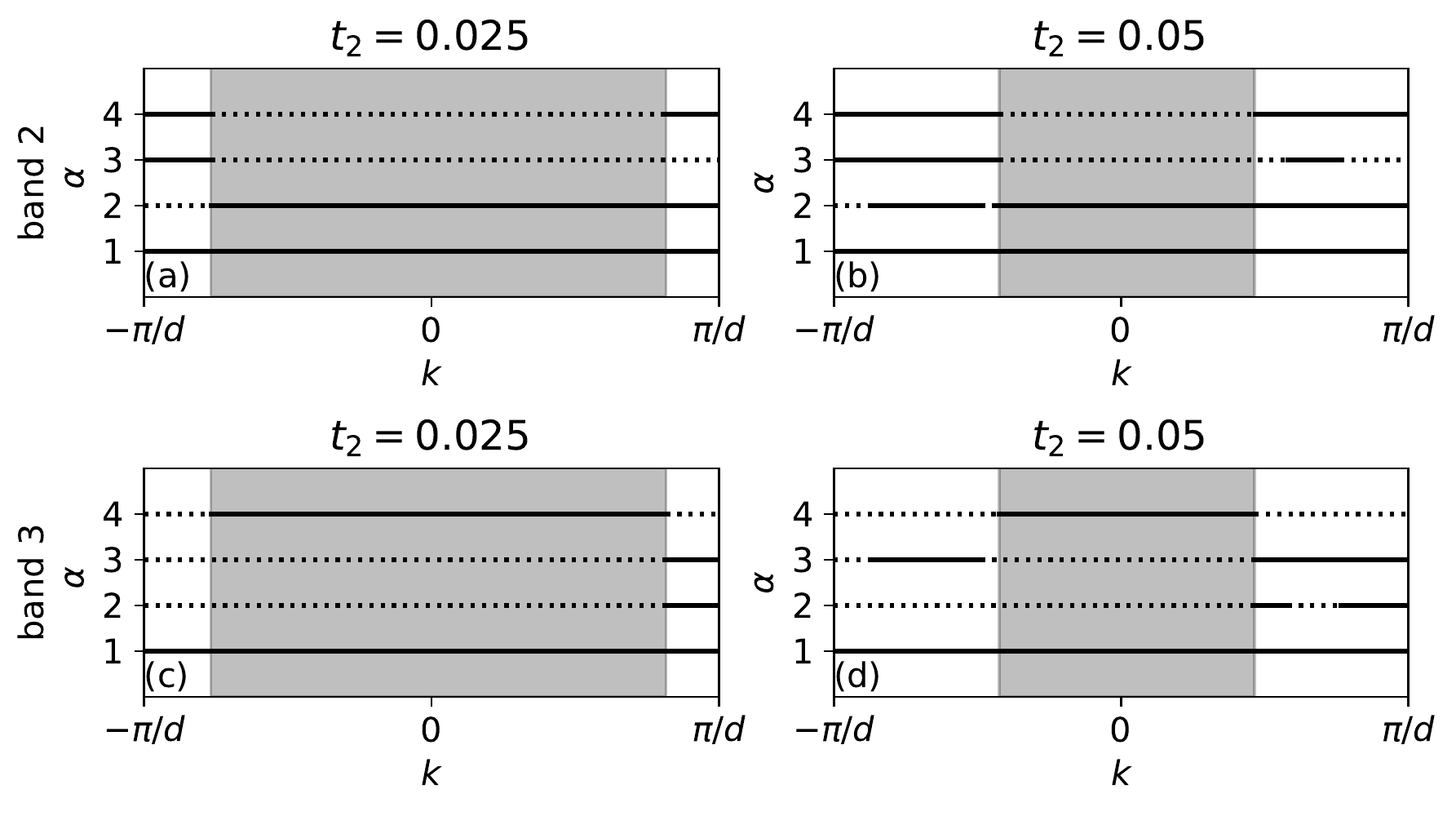}
			\caption{Phases $\phi'_{\alpha,j}$ for band $j = 2$ (a,b) and band $j = 3$ (c,d) for different $t_2$. Solid lines denote  $\phi'_{\alpha,j}=0$, dotted lines  $\phi'_{\alpha,j} = \pi$. Gray shaded areas indicate regions where $\phi'_{\alpha,j}$ is identical to the case $t_2 = 0$ (not shown). These shaded areas are identical to the ones in Fig.\  \ref{fig:phases_Bloch}.}
			\label{fig:phases_Bloch_bands}
		\end{figure}

Instead of phase differences, Fig.\ \ref{fig:phases_Bloch_bands} shows the individual phases $\phi'_{\alpha,j}$ for $j = 2$ (a,b) and $j = 3$ (c,d) in a similar kind of plot. The shaded areas are identical to Fig. \ref{fig:phases_Bloch}. A kind of node-rule can be identified. In the $t_2=0$-like region (shaded area), the values $\phi'_{\alpha,j=2}$ are $0$ for sites $\alpha = 1,2$ and $\pi$ for $\alpha = 3,4$. Hence, the Bloch states $\tilde{u}^{j=2}_\alpha$ are positive for the first two sites and negative for the last two, the state has one ``node''. For band $j = 3$, the Bloch state is positive for sites $\alpha = 1,4$ and negative for $\alpha = 2,3$, i.e., there are two ``nodes''. For $t_2 = 0.025$ the values of $\phi'_{\alpha,j}$ change in the non-shaded area in such a way that band $2$ has two nodes and band $3$ only one, which is indicative of a band inversion. For $t_2 = 0.05$ one can see that a region appears where band $2$ has no nodes and band $3$ even four nodes.
	
		The same gray-shaded areas are shown in the plots of the band structure, Fig. \ref{fig:Bands}. The energy difference between band two and three at the boundaries of the gray shaded area are traced as a function of $t_2$. They are labeled by $\Delta E^-$ (left boundary, negative $k$) and $\Delta E^+$ (right boundary, positive $k$). The energy difference between these bands at $k = 0$ is called $\Delta E^0$. For better visibility, these energies are only indicated in Fig. \ref{fig:Bands}(c,d). Further, the energy differences in the non-shaded areas are always smaller than the energy differences in the gray-shaded areas for $t_2 \lesssim 0.0387$. For $t_2 > 0.0387$, it appears that $\Delta E^0<\Delta E^-$ and certain energy differences in the non-shaded area are larger than differences in the shaded area between the bands $j=2$ and $3$, see Fig. \ref{fig:Bands}(d).  
		
	\subsection{Spectra}\label{sec:HHG}	
		
\begin{figure}
			\begin{minipage}{.99\columnwidth}
				\centering
				\includegraphics[width=\columnwidth]{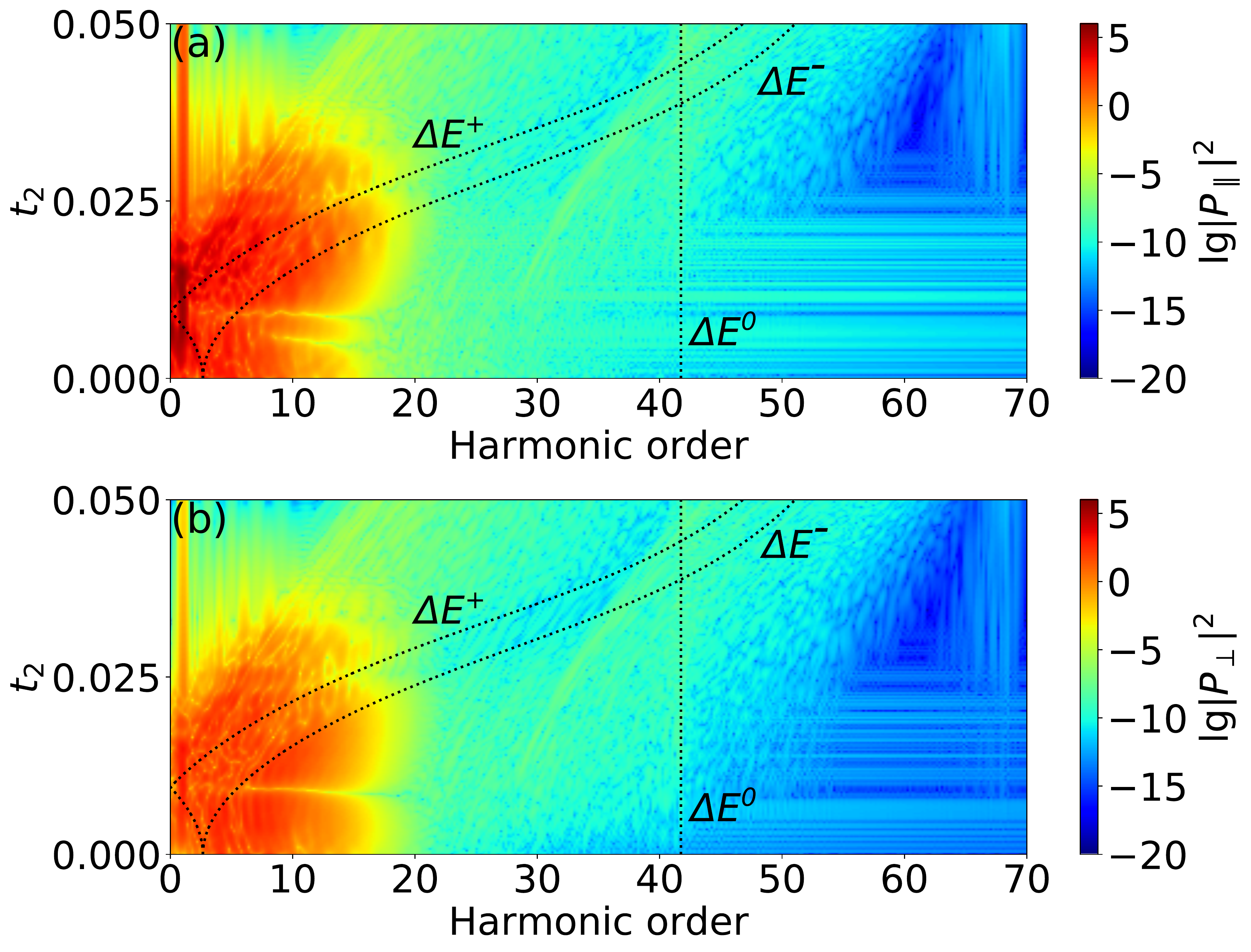}
			\end{minipage}%
			\caption{Harmonic spectra $|P_{\parallel}(\omega)|^2$, $|P_{\perp}(\omega)|^2$  vs $t_2$ in parallel (a) and perpendicular polarization direction (b), respectively. The dotted lines indicate the energy differences defined before (see Fig. \ref{fig:Bands}(c,d)).}
			\label{fig:hhg_haldane}
		\end{figure}

		The harmonic spectra $|P_{\parallel}(\omega)|^2$, $|P_{\perp}(\omega)|^2$  for this system in parallel (a) and perpendicular polarization direction (b), respectively, are shown in Fig. \ref{fig:hhg_haldane} for  $t_2\in[0,0.05]$. The highest yield is found at small harmonic orders and small $t_2$ when the band gap is small compared to the photon energy of the driving field. For larger $t_2$, the band gap increases, and low-order harmonics are due to intraband movement of electrons, which destructively interferes if the valence bands are fully occupied \cite{PhysRevA.96.053418,bauer_high-harmonic_2018}. 
		%Hence, the harmonic yield firstly  decreases with increasing $t_2$.
		As the gap closes at $t_2\simeq 0.01$, the harmonic yield decreases too. This effect appears as a clearly visible horizontal cut in the region of high harmonic yield. 
			
		The three energies $\Delta E^\pm$ and $\Delta E^0$ are indicated in these plots by dotted lines. The energies $\Delta E^\pm$ describe two curves that go through the region of the highest harmonic yield. 
		
\begin{figure}
			\includegraphics[width=\columnwidth]{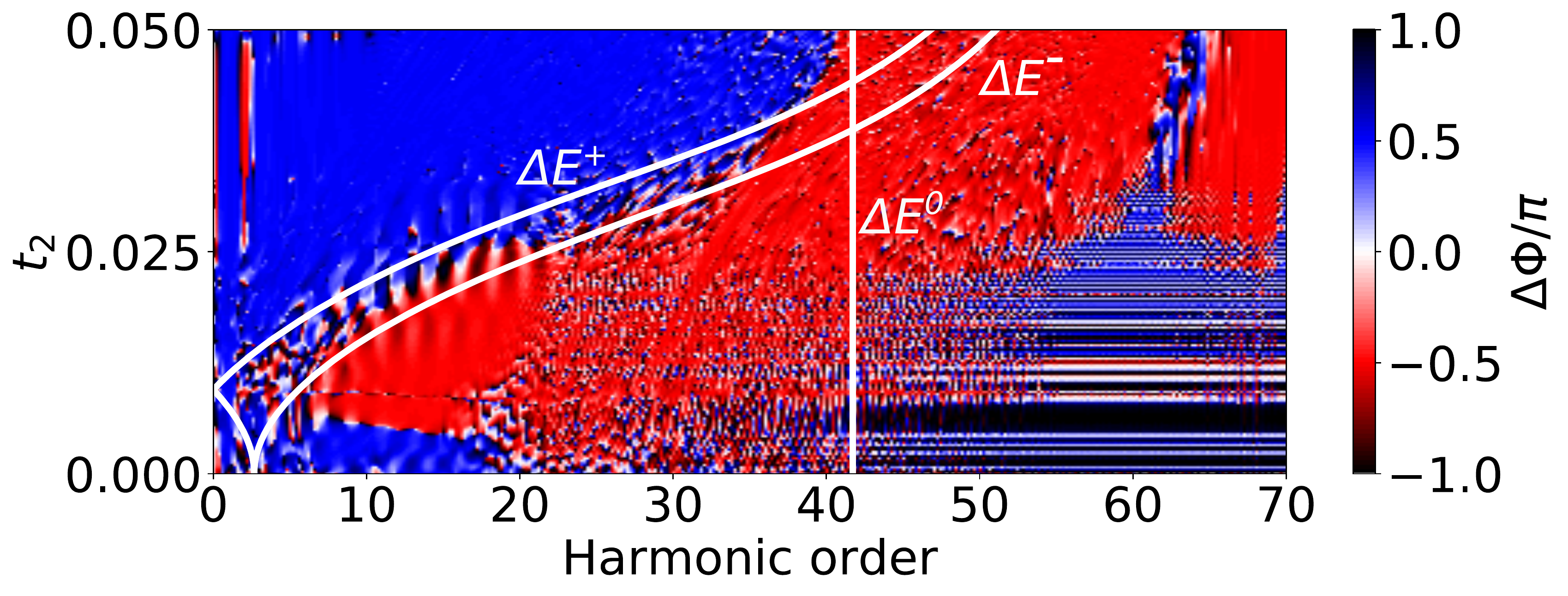}
			\caption{Phase difference (helicity) between the spectra in the two polarization directions.}
			\label{fig:phase_diff}
		\end{figure}

		The emitted light has two polarization directions, the phase difference between those two components can be obtained via equation (\ref{eq:phase}). The phase difference determines the helicity of the emitted photons. The result is shown in Fig. \ref{fig:phase_diff}. The phase difference is preferably $\Delta \Phi = \pm\pi/2$. The interesting fact, and the main result of this work, is that for a given,  sufficiently large $t_2>0.006$ the phase changes from $+\pi/2$ to $-\pi/2$ at a certain harmonic order. This phase flip appears between the energies $\Delta E^\pm$. To be precise, the phase difference is $-\pi/2$ for energies below $\Delta E^+$ and it is $+\pi/2$ for energies larger than $\Delta E^-$, at least as long as $\Delta E^0 > \Delta E^\pm$. 
		
		The case where $\Delta E^0 > \Delta E^\pm$ is examined. As shown previously, the phase difference of the periodic part of the Bloch states $\Delta \phi'_{\alpha}=\phi'_{\alpha,j} - \phi'_{\alpha',j'}$ has a symmetry which changes at certain points in $k$-space for sufficiently large $t_2$. More precise, the phases $\phi'_{\alpha,j = 2}$ and $\phi'_{\alpha',j = 3}$ change at these points in such a way that the properties of band $j = 2$ and $j = 3$ are inverted. 	If the energy difference between these states is smaller than $\Delta E^+$ (non-shaded area in Fig. \ref{fig:Bands}), then $\Delta \phi'_{\alpha}$ is different compared to points where this energy difference is larger than $\Delta E^-$ (gray-shaded area in Fig. \ref{fig:Bands}).
		How $\Delta \phi'_{\alpha}$ affects the current, which finally determines the emitted harmonic radiation,  is shown in Appendix \ref{app:A}.
		The $\Delta E^\pm$ describe the change of the helicity quite well unless $\Delta E^0 < \Delta E^\pm$ (where  $\Delta E^0$ determines the change of the helicity).

	\subsection{Partially filled valence bands}\label{sec:part_filled_bands}
		
\begin{figure}
			\begin{minipage}{.99\columnwidth}
				\centering
				\includegraphics[width=\columnwidth]{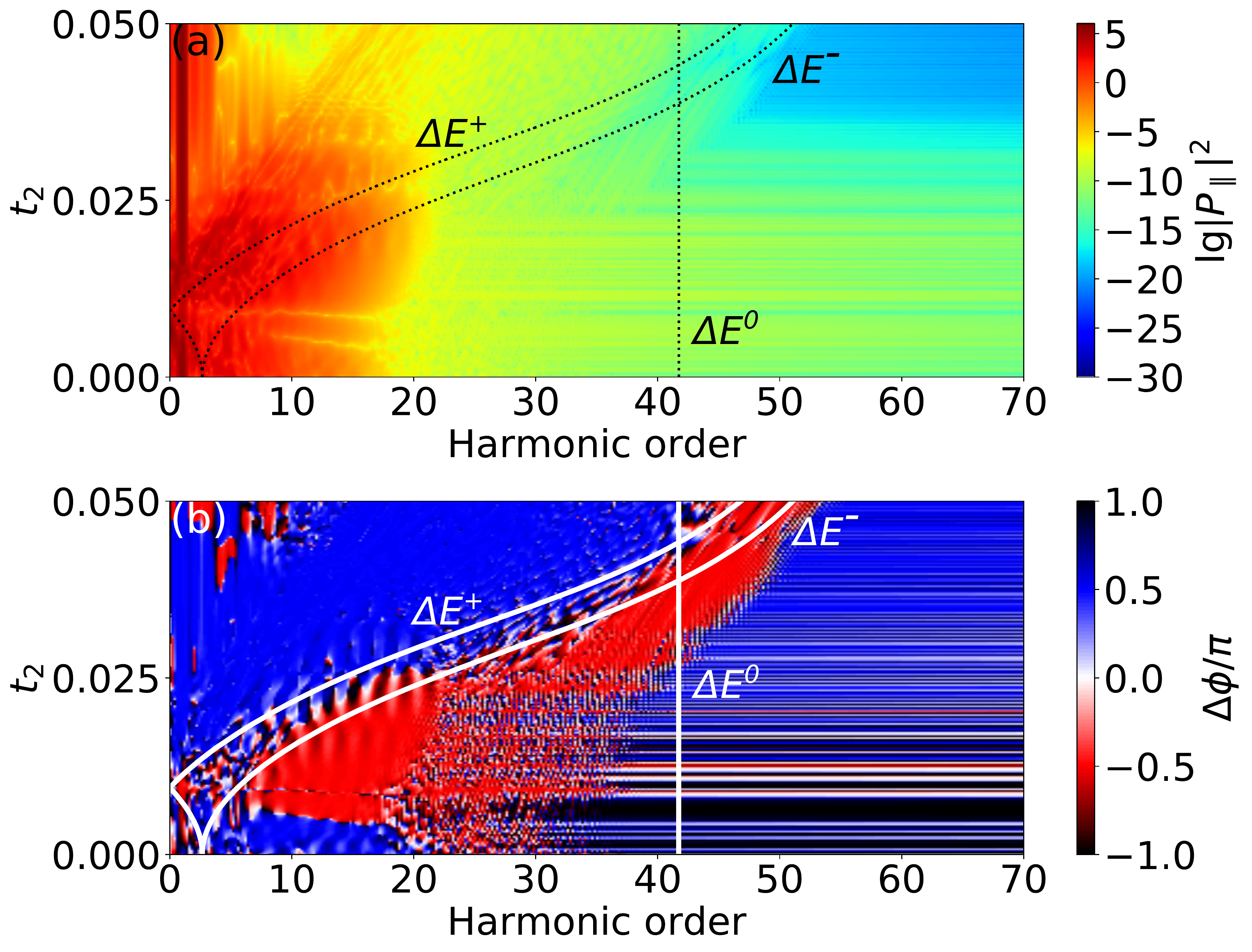}
			\end{minipage}
			\caption{Harmonic spectra $|P_{\parallel}(\omega)|^2$  vs $t_2$ in parallel  polarization direction (a) and helicity (b) calculated from only the ten highest states of the valence bands (i.e., band $j=2$). }
			\label{fig:HHG_50without}
		\end{figure}

		Until now, all the states of the valence bands were occupied, i.e. all states with a negative energy. Now, only the ten highest states of the valence bands are occupied. These states belong to the second valence band $j=2$. In this way, we demonstrate that the electrons occupying the second valence band are responsible for the helicity change of the emitted photons. The result is shown in Fig. \ref{fig:HHG_50without}.
		
		There are only minor differences between the spectra in Fig.\ \ref{fig:hhg_haldane}(a) and Fig.\ \ref{fig:HHG_50without}(a).  
As far as the helicity of the emitted harmonic photons is concerned, one can notice differences between Fig.\ \ref{fig:phase_diff}	 and Fig.\	\ref{fig:HHG_50without}(b) but the phase flip for given $t_2$ still occurs at  the same harmonic order as before.  
	
	\section{Finite nanoribbon with edges}\label{sec:ribbon_edges}
	
		\begin{figure}
			\centering
			\includegraphics[width=\columnwidth]{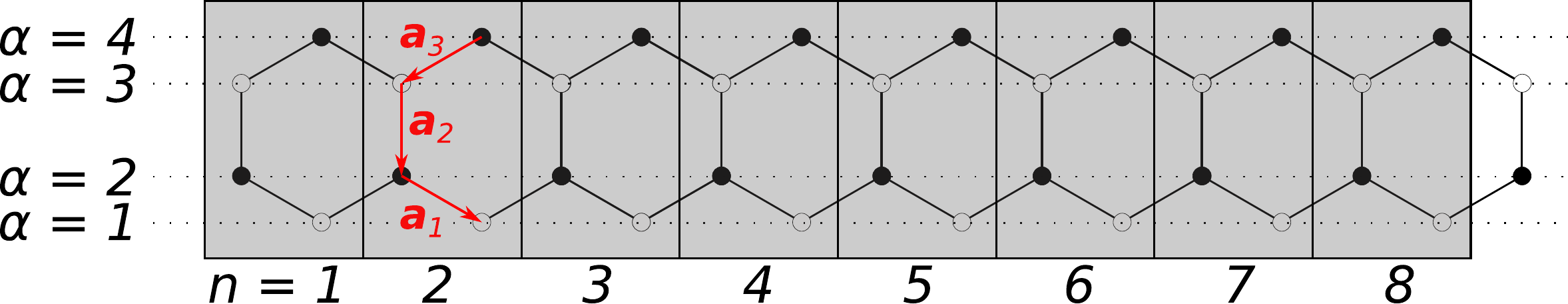}
			\caption{Finite nanoribbon containing $L = 4N+2$ atoms.}
			\label{fig:chain_fin}
		\end{figure}
		
\begin{figure}
			\centering
			\includegraphics[width=\columnwidth]{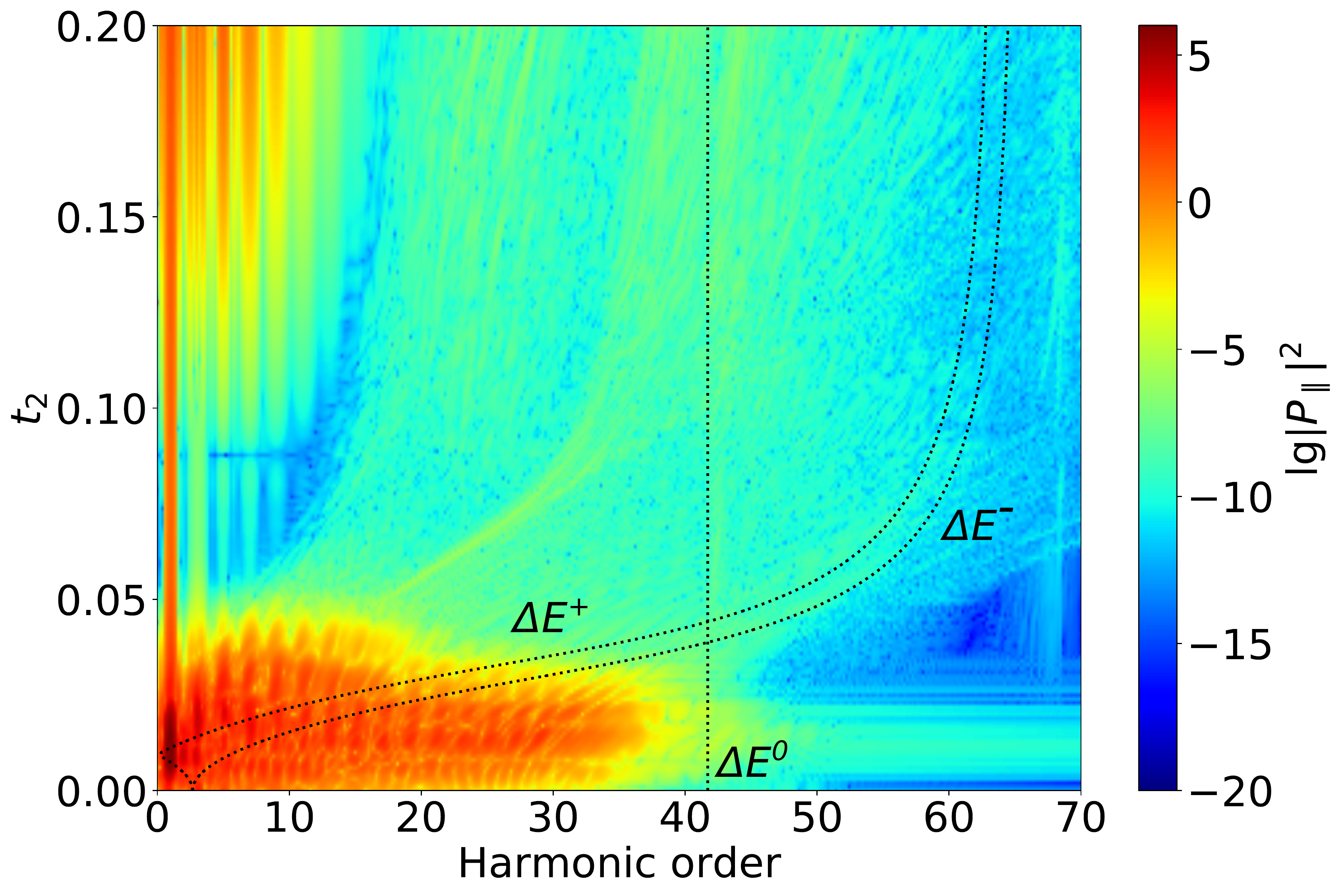}
			\caption{Harmonic spectra $|P_{\parallel}(\omega)|^2$  vs $t_2$ in parallel  polarization direction for the finite system. The dotted lines indicate the energy differences defined by the bulk system, as described before.}
			\label{fig:hhg_haldane_Edge}
		\end{figure}

		In order to show that the helicity flip also occurs in finite nanoribbons despite the fact that the explanation for the flip is based on a bulk analysis, a finite system with edges as shown in Fig. \ref{fig:chain_fin} is investigated. For the ribbon with edges, hopping from the right to the left edge is not possible. The system contains two more atoms compared to the periodic system  in order to complete the hexagon at the right edge.  The cutoff of the plateau for small harmonic orders and small $t_2$ occurs at larger energies as for the periodic system, as seen in Fig. \ref{fig:hhg_haldane_Edge}. 
		
\begin{figure}
			\includegraphics[width=\columnwidth]{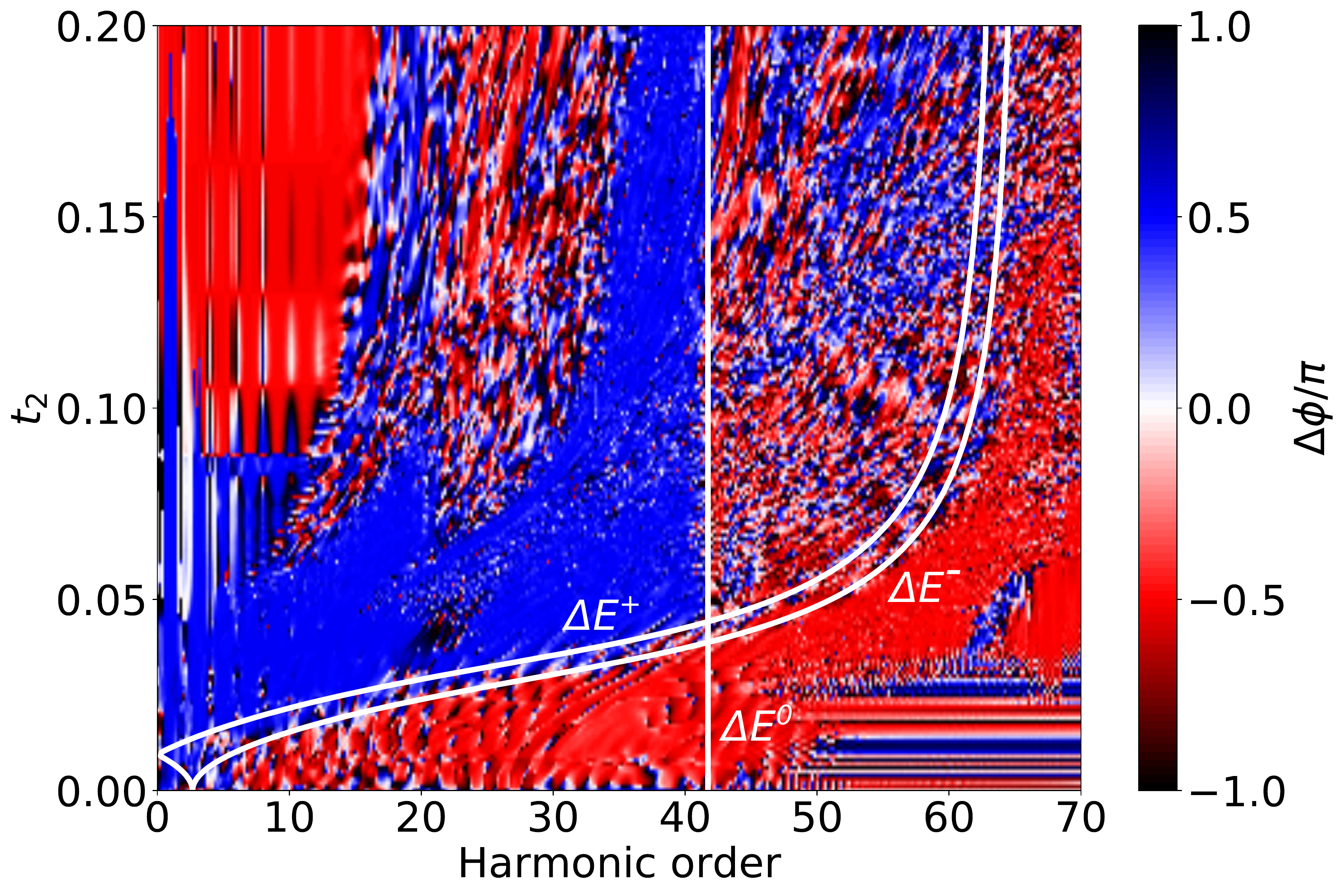}
			\caption{Phase difference (helicity) between the spectra in both polarization directions for the finite ribbon.}
			\label{fig:phase_diff_Edge}
		\end{figure}

		The phase flip discussed for the periodic system can be observed for the finite ribbon as well, see Fig \ref{fig:phase_diff_Edge}. We observe an additional phase flip at around $t_2\approx0.09$ for small odd harmonics (up to order $11$), except for the fundamental (the same flip can be observed for the periodic system, not shown). This helicity flip might be similar to the one for bulk Haldanite explained in \cite{Silva2019}. 
		The authors of \cite{Silva2019} explain their observed phase flip with a topological phase transition in bulk Haldanite. However, there is no topological phase transition around $t_2 =0.09$ in our nanoribbon. Instead, our finite system has edge states, and the energy difference of these states is smallest at $t_2 \approx 0.09$, which coincides with the helicity change. Anyhow, we do not investigate that helicity flip further in this work because the harmonic yield  at $t_2 =0.09$  is very small, and the next-nearest neighbor hopping is almost as big as the nearest neighbor hopping.

	\section{Summary and outlook} \label{sec:summandout}
	
		A helicity flip of the emitted photons in high-harmonic spectra from Haldane-like  nanoribbons  is observed. For a fixed next-nearest neighbor hopping, the helicity of the emitted photons changes at a certain harmonic order. The photon energy where this helicity flip occurs can be predicted by examining the phase differences between the periodic parts of the Bloch states. In previous theoretical works \cite{Silva2019,chacon_observing_2018}, helicity flips of harmonics in bulk Haldanite were observed as a function of the next-nearest neighbor hopping, mapping out the known topological phase transition of the Haldane model.  
		The helicity flip discussed in this work might be observed in realistic systems where the next-nearest neighbor hopping cannot be easily changed. Moreover, the effect might allow to manipulate the helicity of high-harmonic photons, e.g., by multi-color incident laser pulses.

	\section*{Acknowledgment}
	
	C. J. acknowledges financial support by the doctoral fellowship program of the University of Rostock.
	
	\begin{appendix}
	\section{Calculation of the current and the relevance of the phase-differences $\Delta\phi^{j',j}_{\alpha',\alpha}(k_i)$} \label{app:A}

	The current is calculated as the expactation value of the current operator 
	\begin{align}
	\bm{j}^i(t) = \bra{\Psi^i(t)}\hat{\bm{j}}(t)\ket{\Psi^i(t)}
	\end{align}
	with the current operator (\ref{eq:current_operator}). For the state $\ket{\Psi^i(t)}$, one can make a similar Bloch-ansatz as before
	\begin{align}\label{eq:Bloch_ansatz_timedep}
	\ket{\Psi^i(t)} &= \frac{1}{\sqrt{N}}\sum_{m=1}^N \ee^{\ii m k_i d}\ket{m} \otimes \left(u_1(k_i,t)\ee^{\ii k_i d/2}\ket{1} \right.\nonumber\\
	&\left. + u_2(k_i,t)\ket{2} + u_3(k_i,t)\ket{3} + u_4(k_i,t)\ee^{\ii k_i d/2}\ket{4}\right),
	\end{align}
	but now with time-dependent coefficients $u_\alpha(k_i,t)$. 
	Using this ansatz, one ends up with two $4\times 4$ current operators, for the $x$- and $y$-direction. The corresponding currents are then calculated as 
	\begin{align}
	j^{x,y}(k_i,t) = \bm{u}^\dagger(k_i,t) \hat{J}^{x,y}_\mathrm{bulk}(k_i,t)\bm{u}(k_i,t).
	\end{align}
	For the $x$-direction one obtains
	\begin{align}
	\hat{J}^x_\mathrm{bulk}(k_i,t) = \begin{pmatrix}
	f_1(k_i,t) & f_2(k_i,t) & f_3(k_i,t)&0\\
	f_2(k_i,t)&f_4(k_i,t)&0&f_5(k_i,t)\\
	f_3(k_i,t) & 0 & f_1(k_i,t)&f_2(k_i,t)\\
	0&f_5(k_i,t)&f_2(k_i,t)&f_4(k_i,t)
	\end{pmatrix}
	\end{align}
	with 
	\begin{align}
	f_1(k_i,t) &= \ii d t_2\ee^{\ii\phi} \ee^{\ii d (A(t)+k)} + \mathrm{c.c.},\\
	f_2(k_i,t) &= -d t_1\sin \left[(k+A(t)) d /2\right],\\
	f_3(k_i,t) &= \ii\frac{d}{2}t_2\ee^{-\ii\phi}\ee^{\ii d (A(t)+k)/2} + \mathrm{c.c.},\\
	f_4(k_i,t) &= -\ii d t_2\ee^{\ii\phi} \ee^{-\ii d (A(t)+k)} + \mathrm{c.c.},\\
	f_5(k_i,t) &= \ii\frac{d}{2}t_2\ee^{\ii\phi}\ee^{\ii d (A(t)+k)/2} + \mathrm{c.c.},
	\end{align}
	and in $y$-direction
	\begin{align}
	\hat{J}^y_\mathrm{bulk}(k_i,t) = \begin{pmatrix}
	0& g_1(k_i,t) & g_2(k_i,t)&0\\
	-g_1(k_i,t)&0&g_3&g_4(k_i,t)\\
	-g_2(k_i,t) & -g_3 & 0&g_1(k_i,t)\\
	0&-g_4(k_i,t)&-g_1(k_i,t)&0
	\end{pmatrix}
	\end{align}
	with 
	\begin{align}
	g_1(k_i,t) &= \ii a t_1 \cos \left[(k+A(t)) d /2\right],\\
	g_2(k_i,t) &= \ii\frac{3}{2}a t_2\ee^{-\ii\phi}\ee^{\ii d (A(t)+k)/2} -\mathrm{c.c.}, \\
	g_3 &= \ii a t_1,\\
	g_4(k_i,t) &= \ii\frac{3}{2}a t_2\ee^{\ii\phi}\ee^{\ii d (A(t)+k)/2} -\mathrm{c.c.}.
	\end{align}
	Note that all $f_l(k_i,t)$ ($l=1,2,3,4,5$) are real and all $g_s(k_i,t)$ ($s = 1,2,3,4$) are purely imaginary. In the following, the arguments of $f_l(k_i,t)$ and $g_s(k_i,t)$ are dropped. The argument $k_i$ of $u_\alpha(k_i,t)$ is suppressed as well. 
	The current in $x$-direction reads
	\begin{widetext}
		\begin{align}\label{eq:curr_x}
		j^{x}(k_i,t) &= f_1\left(\left|u_1(t)\right|^2 +\left|u_3(t)\right|^2\right) + f_4\left(\left|u_2(t)\right|^2 +\left|u_4(t)\right|^2\right) + f_2\big[u^*_1(t)u_2(t) + u^*_2(t)u_1(t) + u^*_3(t)u_4(t) + u^*_4(t)u_3(t)\big]\nonumber\\ 
		&+ f_3\big[u^*_1(t)u_3(t) + u^*_3(t)u_1(t)\big] + f_5\big[u^*_2(t)u_4(t) + u^*_4(t)u_2(t)\big],
		\end{align}
		and in $y$-direction
		\begin{align}\label{eq:curr_y}
		j^{y}(k_i,t) &= g_1\big[ u^*_1(t)u_2(t) - u^*_2(t)u_1(t) + u^*_3(t)u_4(t) - u^*_4(t)u_3(t)\big]  +  g_2\big[u^*_1(t)u_3(t) - u^*_3(t)u_1(t)\big]\nonumber\\ 
		&+ g_3\big[u^*_2(t)u_3(t) - u^*_3(t)u_2(t)\big] + g_4\big[u^*_2(t)u_4(t) - u^*_4(t)u_2(t)\big].
		\end{align}
	\end{widetext}
	Note that for certain $\alpha$ and $\alpha'$  one has in $x$-direction the terms $u^*_\alpha(t) u_{\alpha'}(t) + \mathrm{c.c.}$ but in  $y$-direction $u^*_\alpha(t) u_{\alpha'}(t) - \mathrm{c.c.}$. The terms in $x$-directions are purely real while in $y$-direction they are purely imaginary. Multiplied with the factors $f_l$ or $g_s$ the current is real in both directions, of course.
	
	For the time-dependent $\bm{u}(k_i,t)$  one should include another index $\bm{u}^j(k_i,t)$, which indicates the occupied band at the start of the laser pulse. For better readability, however, the index of the initial band is dropped in the time dependent functions $\bm{u}(k_i,t)$. 
	Expanding in eigenstates of the unperturbed bulk system,
	\begin{align}
	\bm{u}(k_i,t) = \sum_{j=1}^{4}\tilde{c}_{j}(k_i,t)\bm{u}^j(k_i),
	\end{align} 
	we find
	\begin{align}
	u^*_\alpha(k_i,t)u_{\alpha'}(k_i,t) = \sum_{j,j'=1}^4 \tilde{c}^*_{j}(k_i,t)\tilde{c}_{j'}(k_i,t)u^{j*}_\alpha(k_i)u^{j'}_{\alpha'}(k_i). \label{eq:app1}
	\end{align}
	Writing
	\begin{align}
	u^j_\alpha(k_i) = \left|u^j_\alpha(k_i)\right|\ee^{\ii\phi_{\alpha,j}(k_i)}
	\end{align}
	and
	\begin{align}
	\tilde{c}_j(k_i,t) = \left|\tilde{c}_j(k_i,t)\right|\ee^{\ii\varphi_{j}(k_i,t)},
	\end{align}
	the expression under the sum in \eqref{eq:app1} becomes
	\begin{align}
	&\tilde{c}^*_{j}(k_i,t)\tilde{c}_{j'}(k_i,t)u^{j*}_\alpha(k_i)u^{j'}_{\alpha'}(k_i)\nonumber\\
	&= \tilde{C}_{j',j}(k_i,t)U^{j',j}_{\alpha',\alpha}(k_i)\ee^{\ii\left(\Delta\varphi_{j',j}(k_i,t)+\Delta\phi^{j',j}_{\alpha',\alpha}(k_i)\right)}
	\end{align}
	with $\tilde{C}_{j',j}(k_i,t) = \left|\tilde{c}_{j'}(k_i,t)\right| \left|\tilde{c}_j(k_i,t)\right|$, $U^{j',j}_{\alpha',\alpha}(k_i) = \left|u^{j'}_{\alpha'}(k_i)\right| \left|u^j_\alpha(k_i)\right|$, $\Delta\varphi_{j',j}(k_i,t) = \varphi_{j'}(k_i,t) - \varphi_{j}(k_i,t)$ and $\Delta\phi^{j',j}_{\alpha',\alpha}(k_i) = \phi_{\alpha',j'}(k_i) - \phi_{\alpha,j}(k_i)$.
	\begin{widetext}
		Further, it follows
		\begin{align}
		u^*_\alpha(k_i,t)u_{\alpha'}(k_i,t) + u^*_{\alpha'}(k_i,t)u_{\alpha}(k_i,t) 
		= 2\sum_{j,j'=1}^4 \tilde{C}_{j',j}(k_i,t)U^{j',j}_{\alpha',\alpha}(k_i) \cos \left(\Delta\varphi_{j',j}(k_i,t)+\Delta\phi^{j',j}_{\alpha',\alpha}(k_i)\right)
		\end{align}
		and
		\begin{align}
		u^*_\alpha(k_i,t)u_{\alpha'}(k_i,t) - u^*_{\alpha'}(k_i,t)u_{\alpha}(k_i,t) 
		= 2\ii\sum_{j,j'=1}^4 \tilde{C}_{j',j}(k_i,t)U^{j',j}_{\alpha',\alpha}(k_i) \sin \left(\Delta\varphi_{j',j}(k_i,t)+\Delta\phi^{j',j}_{\alpha',\alpha}(k_i)\right).
		\end{align}
	\end{widetext}
	Hence the expression of the current in $x$-direction (\ref{eq:curr_x}) contains terms  $\sim \cos \left(\Delta\varphi_{j',j}(k_i,t)+\Delta\phi^{j',j}_{\alpha',\alpha}(k_i)\right)$ whereas terms $ \sim\sin \left(\Delta\varphi_{j',j}(k_i,t)+\Delta\phi^{j',j}_{\alpha',\alpha}(k_i)\right)$
	appear in the $y$-component of the current (\ref{eq:curr_y}). 
	
	As it was shown in the main text, the phase-differences $\Delta\phi'_\alpha = \Delta\phi^{3,2}_{\alpha,\alpha}$  change for certain $k_i$. The helicity of the harmonics changes sign at the corresponding harmonic order.		
	The currents depend on these phase-differences, and the dependencies are different for the two directions.
	
	\end{appendix}
	
	\bibliography{biblio.bib}

\begin{thebibliography}{37}%
\makeatletter
\providecommand \@ifxundefined [1]{%
 \@ifx{#1\undefined}
}%
\providecommand \@ifnum [1]{%
 \ifnum #1\expandafter \@firstoftwo
 \else \expandafter \@secondoftwo
 \fi
}%
\providecommand \@ifx [1]{%
 \ifx #1\expandafter \@firstoftwo
 \else \expandafter \@secondoftwo
 \fi
}%
\providecommand \natexlab [1]{#1}%
\providecommand \enquote  [1]{``#1''}%
\providecommand \bibnamefont  [1]{#1}%
\providecommand \bibfnamefont [1]{#1}%
\providecommand \citenamefont [1]{#1}%
\providecommand \href@noop [0]{\@secondoftwo}%
\providecommand \href [0]{\begingroup \@sanitize@url \@href}%
\providecommand \@href[1]{\@@startlink{#1}\@@href}%
\providecommand \@@href[1]{\endgroup#1\@@endlink}%
\providecommand \@sanitize@url [0]{\catcode `\\12\catcode `\$12\catcode
  `\&12\catcode `\#12\catcode `\^12\catcode `\_12\catcode `\%12\relax}%
\providecommand \@@startlink[1]{}%
\providecommand \@@endlink[0]{}%
\providecommand \url  [0]{\begingroup\@sanitize@url \@url }%
\providecommand \@url [1]{\endgroup\@href {#1}{\urlprefix }}%
\providecommand \urlprefix  [0]{URL }%
\providecommand \Eprint [0]{\href }%
\providecommand \doibase [0]{http://dx.doi.org/}%
\providecommand \selectlanguage [0]{\@gobble}%
\providecommand \bibinfo  [0]{\@secondoftwo}%
\providecommand \bibfield  [0]{\@secondoftwo}%
\providecommand \translation [1]{[#1]}%
\providecommand \BibitemOpen [0]{}%
\providecommand \bibitemStop [0]{}%
\providecommand \bibitemNoStop [0]{.\EOS\space}%
\providecommand \EOS [0]{\spacefactor3000\relax}%
\providecommand \BibitemShut  [1]{\csname bibitem#1\endcsname}%
\let\auto@bib@innerbib\@empty
%</preamble>
\bibitem [{\citenamefont {Ghimire}\ \emph {et~al.}(2011)\citenamefont
  {Ghimire}, \citenamefont {DiChiara}, \citenamefont {Sistrunk}, \citenamefont
  {Agostini}, \citenamefont {DiMauro},\ and\ \citenamefont
  {Reis}}]{Ghimire2011}%
  \BibitemOpen
  \bibfield  {author} {\bibinfo {author} {\bibfnamefont {S.}~\bibnamefont
  {Ghimire}}, \bibinfo {author} {\bibfnamefont {A.~D.}\ \bibnamefont
  {DiChiara}}, \bibinfo {author} {\bibfnamefont {E.}~\bibnamefont {Sistrunk}},
  \bibinfo {author} {\bibfnamefont {P.}~\bibnamefont {Agostini}}, \bibinfo
  {author} {\bibfnamefont {L.~F.}\ \bibnamefont {DiMauro}}, \ and\ \bibinfo
  {author} {\bibfnamefont {D.~A.}\ \bibnamefont {Reis}},\ }\href {\doibase
  10.1038/nphys1847} {\bibfield  {journal} {\bibinfo  {journal} {Nat Phys}\
  }\textbf {\bibinfo {volume} {7}},\ \bibinfo {pages} {138} (\bibinfo {year}
  {2011})}\BibitemShut {NoStop}%
\bibitem [{\citenamefont {Schubert}\ \emph {et~al.}(2014)\citenamefont
  {Schubert}, \citenamefont {Hohenleutner}, \citenamefont {Langer},
  \citenamefont {Urbanek}, \citenamefont {Lange}, \citenamefont {Huttner},
  \citenamefont {Golde}, \citenamefont {Meier}, \citenamefont {Kira},
  \citenamefont {Koch},\ and\ \citenamefont {Huber}}]{SchubertO.2014}%
  \BibitemOpen
  \bibfield  {author} {\bibinfo {author} {\bibfnamefont {O.}~\bibnamefont
  {Schubert}}, \bibinfo {author} {\bibfnamefont {M.}~\bibnamefont
  {Hohenleutner}}, \bibinfo {author} {\bibfnamefont {F.}~\bibnamefont
  {Langer}}, \bibinfo {author} {\bibfnamefont {B.}~\bibnamefont {Urbanek}},
  \bibinfo {author} {\bibfnamefont {C.}~\bibnamefont {Lange}}, \bibinfo
  {author} {\bibfnamefont {U.}~\bibnamefont {Huttner}}, \bibinfo {author}
  {\bibfnamefont {D.}~\bibnamefont {Golde}}, \bibinfo {author} {\bibfnamefont
  {T.}~\bibnamefont {Meier}}, \bibinfo {author} {\bibfnamefont
  {M.}~\bibnamefont {Kira}}, \bibinfo {author} {\bibfnamefont {S.}~\bibnamefont
  {Koch}}, \ and\ \bibinfo {author} {\bibfnamefont {R.}~\bibnamefont {Huber}},\
  }\href {http://dx.doi.org/10.1038/nphoton.2013.349} {\bibfield  {journal}
  {\bibinfo  {journal} {Nat Photon}\ }\textbf {\bibinfo {volume} {8}},\
  \bibinfo {pages} {119} (\bibinfo {year} {2014})},\ \bibinfo {note}
  {letter}\BibitemShut {NoStop}%
\bibitem [{\citenamefont {Vampa}\ \emph {et~al.}(2015)\citenamefont {Vampa},
  \citenamefont {Hammond}, \citenamefont {Thir\'e}, \citenamefont {Schmidt},
  \citenamefont {L\'egar\'e}, \citenamefont {McDonald}, \citenamefont {Brabec},
  \citenamefont {Klug},\ and\ \citenamefont
  {Corkum}}]{VampaPhysRevLett.115.193603}%
  \BibitemOpen
  \bibfield  {author} {\bibinfo {author} {\bibfnamefont {G.}~\bibnamefont
  {Vampa}}, \bibinfo {author} {\bibfnamefont {T.~J.}\ \bibnamefont {Hammond}},
  \bibinfo {author} {\bibfnamefont {N.}~\bibnamefont {Thir\'e}}, \bibinfo
  {author} {\bibfnamefont {B.~E.}\ \bibnamefont {Schmidt}}, \bibinfo {author}
  {\bibfnamefont {F.}~\bibnamefont {L\'egar\'e}}, \bibinfo {author}
  {\bibfnamefont {C.~R.}\ \bibnamefont {McDonald}}, \bibinfo {author}
  {\bibfnamefont {T.}~\bibnamefont {Brabec}}, \bibinfo {author} {\bibfnamefont
  {D.~D.}\ \bibnamefont {Klug}}, \ and\ \bibinfo {author} {\bibfnamefont
  {P.~B.}\ \bibnamefont {Corkum}},\ }\href {\doibase
  10.1103/PhysRevLett.115.193603} {\bibfield  {journal} {\bibinfo  {journal}
  {Phys. Rev. Lett.}\ }\textbf {\bibinfo {volume} {115}},\ \bibinfo {pages}
  {193603} (\bibinfo {year} {2015})}\BibitemShut {NoStop}%
\bibitem [{\citenamefont {Hohenleutner}\ \emph {et~al.}(2015)\citenamefont
  {Hohenleutner}, \citenamefont {Langer}, \citenamefont {Schubert},
  \citenamefont {Knorr}, \citenamefont {Huttner}, \citenamefont {Koch},
  \citenamefont {Kira},\ and\ \citenamefont {Huber}}]{Hohenleutner2015}%
  \BibitemOpen
  \bibfield  {author} {\bibinfo {author} {\bibfnamefont {M.}~\bibnamefont
  {Hohenleutner}}, \bibinfo {author} {\bibfnamefont {F.}~\bibnamefont
  {Langer}}, \bibinfo {author} {\bibfnamefont {O.}~\bibnamefont {Schubert}},
  \bibinfo {author} {\bibfnamefont {M.}~\bibnamefont {Knorr}}, \bibinfo
  {author} {\bibfnamefont {U.}~\bibnamefont {Huttner}}, \bibinfo {author}
  {\bibfnamefont {S.~W.}\ \bibnamefont {Koch}}, \bibinfo {author}
  {\bibfnamefont {M.}~\bibnamefont {Kira}}, \ and\ \bibinfo {author}
  {\bibfnamefont {R.}~\bibnamefont {Huber}},\ }\href
  {http://dx.doi.org/10.1038/nature14652} {\bibfield  {journal} {\bibinfo
  {journal} {Nature}\ }\textbf {\bibinfo {volume} {523}},\ \bibinfo {pages}
  {572} (\bibinfo {year} {2015})},\ \bibinfo {note} {letter}\BibitemShut
  {NoStop}%
\bibitem [{\citenamefont {Luu}\ \emph {et~al.}(2015)\citenamefont {Luu},
  \citenamefont {Garg}, \citenamefont {Kruchinin}, \citenamefont {Moulet},
  \citenamefont {Hassan},\ and\ \citenamefont {Goulielmakis}}]{Luu2015}%
  \BibitemOpen
  \bibfield  {author} {\bibinfo {author} {\bibfnamefont {T.~T.}\ \bibnamefont
  {Luu}}, \bibinfo {author} {\bibfnamefont {M.}~\bibnamefont {Garg}}, \bibinfo
  {author} {\bibfnamefont {S.~Y.}\ \bibnamefont {Kruchinin}}, \bibinfo {author}
  {\bibfnamefont {A.}~\bibnamefont {Moulet}}, \bibinfo {author} {\bibfnamefont
  {M.~T.}\ \bibnamefont {Hassan}}, \ and\ \bibinfo {author} {\bibfnamefont
  {E.}~\bibnamefont {Goulielmakis}},\ }\href
  {http://dx.doi.org/10.1038/nature14456} {\bibfield  {journal} {\bibinfo
  {journal} {Nature}\ }\textbf {\bibinfo {volume} {521}},\ \bibinfo {pages}
  {498} (\bibinfo {year} {2015})},\ \bibinfo {note} {letter}\BibitemShut
  {NoStop}%
\bibitem [{\citenamefont {Ndabashimiye}\ \emph {et~al.}()\citenamefont
  {Ndabashimiye}, \citenamefont {Ghimire}, \citenamefont {Wu}, \citenamefont
  {Browne}, \citenamefont {Schafer}, \citenamefont {Gaarde},\ and\
  \citenamefont {Reis}}]{ndabashimiye_solid-state_2016}%
  \BibitemOpen
  \bibfield  {author} {\bibinfo {author} {\bibfnamefont {G.}~\bibnamefont
  {Ndabashimiye}}, \bibinfo {author} {\bibfnamefont {S.}~\bibnamefont
  {Ghimire}}, \bibinfo {author} {\bibfnamefont {M.}~\bibnamefont {Wu}},
  \bibinfo {author} {\bibfnamefont {D.~A.}\ \bibnamefont {Browne}}, \bibinfo
  {author} {\bibfnamefont {K.~J.}\ \bibnamefont {Schafer}}, \bibinfo {author}
  {\bibfnamefont {M.~B.}\ \bibnamefont {Gaarde}}, \ and\ \bibinfo {author}
  {\bibfnamefont {D.~A.}\ \bibnamefont {Reis}},\ }\href {\doibase
  10.1038/nature17660} {\bibfield  {journal} {\bibinfo  {journal} {Nature}\
  }\textbf {\bibinfo {volume} {534}},\ \bibinfo {pages} {520}}\BibitemShut
  {NoStop}%
\bibitem [{\citenamefont {Langer}\ \emph {et~al.}(2017)\citenamefont {Langer},
  \citenamefont {Hohenleutner}, \citenamefont {Huttner}, \citenamefont {Koch},
  \citenamefont {Kira},\ and\ \citenamefont {Huber}}]{LangerF.2017}%
  \BibitemOpen
  \bibfield  {author} {\bibinfo {author} {\bibfnamefont {F.}~\bibnamefont
  {Langer}}, \bibinfo {author} {\bibfnamefont {M.}~\bibnamefont
  {Hohenleutner}}, \bibinfo {author} {\bibfnamefont {U.}~\bibnamefont
  {Huttner}}, \bibinfo {author} {\bibfnamefont {S.}~\bibnamefont {Koch}},
  \bibinfo {author} {\bibfnamefont {M.}~\bibnamefont {Kira}}, \ and\ \bibinfo
  {author} {\bibfnamefont {R.}~\bibnamefont {Huber}},\ }\href
  {http://dx.doi.org/10.1038/nphoton.2017.29} {\bibfield  {journal} {\bibinfo
  {journal} {Nat Photon}\ }\textbf {\bibinfo {volume} {11}},\ \bibinfo {pages}
  {227} (\bibinfo {year} {2017})},\ \bibinfo {note} {letter}\BibitemShut
  {NoStop}%
\bibitem [{\citenamefont {Tancogne-Dejean}\ \emph {et~al.}(2017)\citenamefont
  {Tancogne-Dejean}, \citenamefont {M\"ucke}, \citenamefont {K\"artner},\ and\
  \citenamefont {Rubio}}]{TancPhysRevLett.118.087403}%
  \BibitemOpen
  \bibfield  {author} {\bibinfo {author} {\bibfnamefont {N.}~\bibnamefont
  {Tancogne-Dejean}}, \bibinfo {author} {\bibfnamefont {O.~D.}\ \bibnamefont
  {M\"ucke}}, \bibinfo {author} {\bibfnamefont {F.~X.}\ \bibnamefont
  {K\"artner}}, \ and\ \bibinfo {author} {\bibfnamefont {A.}~\bibnamefont
  {Rubio}},\ }\href {\doibase 10.1103/PhysRevLett.118.087403} {\bibfield
  {journal} {\bibinfo  {journal} {Phys. Rev. Lett.}\ }\textbf {\bibinfo
  {volume} {118}},\ \bibinfo {pages} {087403} (\bibinfo {year}
  {2017})}\BibitemShut {NoStop}%
\bibitem [{\citenamefont {You}\ \emph {et~al.}(2017)\citenamefont {You},
  \citenamefont {Yin}, \citenamefont {Wu}, \citenamefont {Chew}, \citenamefont
  {Ren}, \citenamefont {Zhuang}, \citenamefont {Gholam-Mirzaei}, \citenamefont
  {Chini}, \citenamefont {Chang},\ and\ \citenamefont
  {Ghimire}}]{you_high-harmonic_2017}%
  \BibitemOpen
  \bibfield  {author} {\bibinfo {author} {\bibfnamefont {Y.~S.}\ \bibnamefont
  {You}}, \bibinfo {author} {\bibfnamefont {Y.}~\bibnamefont {Yin}}, \bibinfo
  {author} {\bibfnamefont {Y.}~\bibnamefont {Wu}}, \bibinfo {author}
  {\bibfnamefont {A.}~\bibnamefont {Chew}}, \bibinfo {author} {\bibfnamefont
  {X.}~\bibnamefont {Ren}}, \bibinfo {author} {\bibfnamefont {F.}~\bibnamefont
  {Zhuang}}, \bibinfo {author} {\bibfnamefont {S.}~\bibnamefont
  {Gholam-Mirzaei}}, \bibinfo {author} {\bibfnamefont {M.}~\bibnamefont
  {Chini}}, \bibinfo {author} {\bibfnamefont {Z.}~\bibnamefont {Chang}}, \ and\
  \bibinfo {author} {\bibfnamefont {S.}~\bibnamefont {Ghimire}},\ }\href
  {\doibase 10.1038/s41467-017-00989-4} {\bibfield  {journal} {\bibinfo
  {journal} {Nature Communications}\ }\textbf {\bibinfo {volume} {8}},\
  \bibinfo {pages} {724} (\bibinfo {year} {2017})}\BibitemShut {NoStop}%
\bibitem [{\citenamefont {Zhang}\ \emph {et~al.}(2018)\citenamefont {Zhang},
  \citenamefont {Si}, \citenamefont {Murakami}, \citenamefont {Bai},\ and\
  \citenamefont {George}}]{Zhang2018}%
  \BibitemOpen
  \bibfield  {author} {\bibinfo {author} {\bibfnamefont {G.~P.}\ \bibnamefont
  {Zhang}}, \bibinfo {author} {\bibfnamefont {M.~S.}\ \bibnamefont {Si}},
  \bibinfo {author} {\bibfnamefont {M.}~\bibnamefont {Murakami}}, \bibinfo
  {author} {\bibfnamefont {Y.~H.}\ \bibnamefont {Bai}}, \ and\ \bibinfo
  {author} {\bibfnamefont {T.~F.}\ \bibnamefont {George}},\ }\href {\doibase
  10.1038/s41467-018-05535-4} {\bibfield  {journal} {\bibinfo  {journal}
  {Nature Communications}\ }\textbf {\bibinfo {volume} {9}},\ \bibinfo {pages}
  {3031} (\bibinfo {year} {2018})}\BibitemShut {NoStop}%
\bibitem [{\citenamefont {Vampa}\ \emph {et~al.}(2018)\citenamefont {Vampa},
  \citenamefont {Hammond}, \citenamefont {Taucer}, \citenamefont {Ding},
  \citenamefont {Ropagnol}, \citenamefont {Ozaki}, \citenamefont {Delprat},
  \citenamefont {Chaker}, \citenamefont {Thir{\'e}}, \citenamefont {Schmidt},
  \citenamefont {L{\'e}gar{\'e}}, \citenamefont {Klug}, \citenamefont {Naumov},
  \citenamefont {Villeneuve}, \citenamefont {Staudte},\ and\ \citenamefont
  {Corkum}}]{Vampa2018}%
  \BibitemOpen
  \bibfield  {author} {\bibinfo {author} {\bibfnamefont {G.}~\bibnamefont
  {Vampa}}, \bibinfo {author} {\bibfnamefont {T.~J.}\ \bibnamefont {Hammond}},
  \bibinfo {author} {\bibfnamefont {M.}~\bibnamefont {Taucer}}, \bibinfo
  {author} {\bibfnamefont {X.}~\bibnamefont {Ding}}, \bibinfo {author}
  {\bibfnamefont {X.}~\bibnamefont {Ropagnol}}, \bibinfo {author}
  {\bibfnamefont {T.}~\bibnamefont {Ozaki}}, \bibinfo {author} {\bibfnamefont
  {S.}~\bibnamefont {Delprat}}, \bibinfo {author} {\bibfnamefont
  {M.}~\bibnamefont {Chaker}}, \bibinfo {author} {\bibfnamefont
  {N.}~\bibnamefont {Thir{\'e}}}, \bibinfo {author} {\bibfnamefont {B.~E.}\
  \bibnamefont {Schmidt}}, \bibinfo {author} {\bibfnamefont {F.}~\bibnamefont
  {L{\'e}gar{\'e}}}, \bibinfo {author} {\bibfnamefont {D.~D.}\ \bibnamefont
  {Klug}}, \bibinfo {author} {\bibfnamefont {A.~Y.}\ \bibnamefont {Naumov}},
  \bibinfo {author} {\bibfnamefont {D.~M.}\ \bibnamefont {Villeneuve}},
  \bibinfo {author} {\bibfnamefont {A.}~\bibnamefont {Staudte}}, \ and\
  \bibinfo {author} {\bibfnamefont {P.~B.}\ \bibnamefont {Corkum}},\ }\href
  {\doibase 10.1038/s41566-018-0193-5} {\bibfield  {journal} {\bibinfo
  {journal} {Nature Photonics}\ }\textbf {\bibinfo {volume} {12}},\ \bibinfo
  {pages} {465} (\bibinfo {year} {2018})}\BibitemShut {NoStop}%
\bibitem [{\citenamefont {Baudisch}\ \emph {et~al.}(2018)\citenamefont
  {Baudisch}, \citenamefont {Marini}, \citenamefont {Cox}, \citenamefont {Zhu},
  \citenamefont {Silva}, \citenamefont {Teichmann}, \citenamefont {Massicotte},
  \citenamefont {Koppens}, \citenamefont {Levitov}, \citenamefont
  {Garc{\'i}a~de Abajo},\ and\ \citenamefont {Biegert}}]{Baudisch2018}%
  \BibitemOpen
  \bibfield  {author} {\bibinfo {author} {\bibfnamefont {M.}~\bibnamefont
  {Baudisch}}, \bibinfo {author} {\bibfnamefont {A.}~\bibnamefont {Marini}},
  \bibinfo {author} {\bibfnamefont {J.~D.}\ \bibnamefont {Cox}}, \bibinfo
  {author} {\bibfnamefont {T.}~\bibnamefont {Zhu}}, \bibinfo {author}
  {\bibfnamefont {F.}~\bibnamefont {Silva}}, \bibinfo {author} {\bibfnamefont
  {S.}~\bibnamefont {Teichmann}}, \bibinfo {author} {\bibfnamefont
  {M.}~\bibnamefont {Massicotte}}, \bibinfo {author} {\bibfnamefont
  {F.}~\bibnamefont {Koppens}}, \bibinfo {author} {\bibfnamefont {L.~S.}\
  \bibnamefont {Levitov}}, \bibinfo {author} {\bibfnamefont {F.~J.}\
  \bibnamefont {Garc{\'i}a~de Abajo}}, \ and\ \bibinfo {author} {\bibfnamefont
  {J.}~\bibnamefont {Biegert}},\ }\href {\doibase 10.1038/s41467-018-03413-7}
  {\bibfield  {journal} {\bibinfo  {journal} {Nature Communications}\ }\textbf
  {\bibinfo {volume} {9}},\ \bibinfo {pages} {1018} (\bibinfo {year}
  {2018})}\BibitemShut {NoStop}%
\bibitem [{\citenamefont {Garg}\ \emph {et~al.}(2018)\citenamefont {Garg},
  \citenamefont {Kim},\ and\ \citenamefont {Goulielmakis}}]{Garg2018}%
  \BibitemOpen
  \bibfield  {author} {\bibinfo {author} {\bibfnamefont {M.}~\bibnamefont
  {Garg}}, \bibinfo {author} {\bibfnamefont {H.~Y.}\ \bibnamefont {Kim}}, \
  and\ \bibinfo {author} {\bibfnamefont {E.}~\bibnamefont {Goulielmakis}},\
  }\href {\doibase 10.1038/s41566-018-0123-6} {\bibfield  {journal} {\bibinfo
  {journal} {Nature Photonics}\ }\textbf {\bibinfo {volume} {12}},\ \bibinfo
  {pages} {291} (\bibinfo {year} {2018})}\BibitemShut {NoStop}%
\bibitem [{\citenamefont {Abadie}\ \emph {et~al.}(2018)\citenamefont {Abadie},
  \citenamefont {Wu},\ and\ \citenamefont {Gaarde}}]{Abadie2018}%
  \BibitemOpen
  \bibfield  {author} {\bibinfo {author} {\bibfnamefont {C.~Q.}\ \bibnamefont
  {Abadie}}, \bibinfo {author} {\bibfnamefont {M.}~\bibnamefont {Wu}}, \ and\
  \bibinfo {author} {\bibfnamefont {M.~B.}\ \bibnamefont {Gaarde}},\ }\href
  {\doibase 10.1364/OL.43.005339} {\bibfield  {journal} {\bibinfo  {journal}
  {Opt. Lett.}\ }\textbf {\bibinfo {volume} {43}},\ \bibinfo {pages} {5339}
  (\bibinfo {year} {2018})}\BibitemShut {NoStop}%
\bibitem [{\citenamefont {Yue}\ and\ \citenamefont {Gaarde}(2020)}]{Yue2020}%
  \BibitemOpen
  \bibfield  {author} {\bibinfo {author} {\bibfnamefont {L.}~\bibnamefont
  {Yue}}\ and\ \bibinfo {author} {\bibfnamefont {M.~B.}\ \bibnamefont
  {Gaarde}},\ }\href {\doibase 10.1103/PhysRevA.101.053411} {\bibfield
  {journal} {\bibinfo  {journal} {Phys. Rev. A}\ }\textbf {\bibinfo {volume}
  {101}},\ \bibinfo {pages} {053411} (\bibinfo {year} {2020})}\BibitemShut
  {NoStop}%
\bibitem [{\citenamefont {Hasan}\ and\ \citenamefont
  {Kane}(2010)}]{topinsRevModPhys.82.3045}%
  \BibitemOpen
  \bibfield  {author} {\bibinfo {author} {\bibfnamefont {M.~Z.}\ \bibnamefont
  {Hasan}}\ and\ \bibinfo {author} {\bibfnamefont {C.~L.}\ \bibnamefont
  {Kane}},\ }\href {\doibase 10.1103/RevModPhys.82.3045} {\bibfield  {journal}
  {\bibinfo  {journal} {Rev. Mod. Phys.}\ }\textbf {\bibinfo {volume} {82}},\
  \bibinfo {pages} {3045} (\bibinfo {year} {2010})}\BibitemShut {NoStop}%
\bibitem [{\citenamefont {Franz}\ and\ \citenamefont
  {Molenkamp}(2013)}]{topins}%
  \BibitemOpen
  \bibinfo {editor} {\bibfnamefont {M.}~\bibnamefont {Franz}}\ and\ \bibinfo
  {editor} {\bibfnamefont {L.}~\bibnamefont {Molenkamp}},\ eds.,\ \href
  {\doibase https://doi.org/10.1016/B978-0-444-63314-9.00012-3} {\emph
  {\bibinfo {title} {Topological Insulators}}},\ \bibinfo {series}
  {Contemporary Concepts of Condensed Matter Science}, Vol.~\bibinfo {volume}
  {6}\ (\bibinfo  {publisher} {Elsevier},\ \bibinfo {year} {2013})\BibitemShut
  {NoStop}%
\bibitem [{\citenamefont {Asb{\'o}th}\ \emph {et~al.}(2016)\citenamefont
  {Asb{\'o}th}, \citenamefont {Oroszl{\'a}ny},\ and\ \citenamefont
  {P\'alyi}}]{topinsshortcourse}%
  \BibitemOpen
  \bibfield  {author} {\bibinfo {author} {\bibfnamefont {J.}~\bibnamefont
  {Asb{\'o}th}}, \bibinfo {author} {\bibfnamefont {L.}~\bibnamefont
  {Oroszl{\'a}ny}}, \ and\ \bibinfo {author} {\bibfnamefont {A.}~\bibnamefont
  {P\'alyi}},\ }\href@noop {} {\emph {\bibinfo {title} {A Short Course on
  Topological Insulators}}},\ \bibinfo {series} {Lecture Notes in Physics},
  Vol.\ \bibinfo {volume} {919}\ (\bibinfo  {publisher} {Springer},\ \bibinfo
  {year} {2016})\BibitemShut {NoStop}%
\bibitem [{\citenamefont {Reimann}\ \emph {et~al.}(2018)\citenamefont
  {Reimann}, \citenamefont {Schlauderer}, \citenamefont {Schmid}, \citenamefont
  {Langer}, \citenamefont {Baierl}, \citenamefont {Kokh}, \citenamefont
  {Tereshchenko}, \citenamefont {Kimura}, \citenamefont {Lange}, \citenamefont
  {G{\"u}dde}, \citenamefont {H{\"o}fer},\ and\ \citenamefont
  {Huber}}]{Reimann2018}%
  \BibitemOpen
  \bibfield  {author} {\bibinfo {author} {\bibfnamefont {J.}~\bibnamefont
  {Reimann}}, \bibinfo {author} {\bibfnamefont {S.}~\bibnamefont
  {Schlauderer}}, \bibinfo {author} {\bibfnamefont {C.~P.}\ \bibnamefont
  {Schmid}}, \bibinfo {author} {\bibfnamefont {F.}~\bibnamefont {Langer}},
  \bibinfo {author} {\bibfnamefont {S.}~\bibnamefont {Baierl}}, \bibinfo
  {author} {\bibfnamefont {K.~A.}\ \bibnamefont {Kokh}}, \bibinfo {author}
  {\bibfnamefont {O.~E.}\ \bibnamefont {Tereshchenko}}, \bibinfo {author}
  {\bibfnamefont {A.}~\bibnamefont {Kimura}}, \bibinfo {author} {\bibfnamefont
  {C.}~\bibnamefont {Lange}}, \bibinfo {author} {\bibfnamefont
  {J.}~\bibnamefont {G{\"u}dde}}, \bibinfo {author} {\bibfnamefont
  {U.}~\bibnamefont {H{\"o}fer}}, \ and\ \bibinfo {author} {\bibfnamefont
  {R.}~\bibnamefont {Huber}},\ }\href {\doibase 10.1038/s41586-018-0544-x}
  {\bibfield  {journal} {\bibinfo  {journal} {Nature}\ }\textbf {\bibinfo
  {volume} {562}},\ \bibinfo {pages} {396} (\bibinfo {year}
  {2018})}\BibitemShut {NoStop}%
\bibitem [{\citenamefont {Koochaki~Kelardeh}\ \emph {et~al.}(2017)\citenamefont
  {Koochaki~Kelardeh}, \citenamefont {Apalkov},\ and\ \citenamefont
  {Stockman}}]{PhysRevB.96.075409}%
  \BibitemOpen
  \bibfield  {author} {\bibinfo {author} {\bibfnamefont {H.}~\bibnamefont
  {Koochaki~Kelardeh}}, \bibinfo {author} {\bibfnamefont {V.}~\bibnamefont
  {Apalkov}}, \ and\ \bibinfo {author} {\bibfnamefont {M.~I.}\ \bibnamefont
  {Stockman}},\ }\href {\doibase 10.1103/PhysRevB.96.075409} {\bibfield
  {journal} {\bibinfo  {journal} {Phys. Rev. B}\ }\textbf {\bibinfo {volume}
  {96}},\ \bibinfo {pages} {075409} (\bibinfo {year} {2017})}\BibitemShut
  {NoStop}%
\bibitem [{\citenamefont {Bauer}\ and\ \citenamefont
  {Hansen}(2018)}]{bauer_high-harmonic_2018}%
  \BibitemOpen
  \bibfield  {author} {\bibinfo {author} {\bibfnamefont {D.}~\bibnamefont
  {Bauer}}\ and\ \bibinfo {author} {\bibfnamefont {K.~K.}\ \bibnamefont
  {Hansen}},\ }\href {\doibase 10.1103/PhysRevLett.120.177401} {\bibfield
  {journal} {\bibinfo  {journal} {Phys. Rev. Lett.}\ }\textbf {\bibinfo
  {volume} {120}},\ \bibinfo {pages} {177401} (\bibinfo {year}
  {2018})}\BibitemShut {NoStop}%
\bibitem [{\citenamefont {Silva}\ \emph {et~al.}(2019)\citenamefont {Silva},
  \citenamefont {Jim{\'e}nez-Gal{\'a}n}, \citenamefont {Amorim}, \citenamefont
  {Smirnova},\ and\ \citenamefont {Ivanov}}]{Silva2019}%
  \BibitemOpen
  \bibfield  {author} {\bibinfo {author} {\bibfnamefont {R.~E.~F.}\
  \bibnamefont {Silva}}, \bibinfo {author} {\bibfnamefont {{\'A}.}~\bibnamefont
  {Jim{\'e}nez-Gal{\'a}n}}, \bibinfo {author} {\bibfnamefont {B.}~\bibnamefont
  {Amorim}}, \bibinfo {author} {\bibfnamefont {O.}~\bibnamefont {Smirnova}}, \
  and\ \bibinfo {author} {\bibfnamefont {M.}~\bibnamefont {Ivanov}},\ }\href
  {\doibase 10.1038/s41566-019-0516-1} {\bibfield  {journal} {\bibinfo
  {journal} {Nature Photonics}\ }\textbf {\bibinfo {volume} {13}},\ \bibinfo
  {pages} {849} (\bibinfo {year} {2019})}\BibitemShut {NoStop}%
\bibitem [{\citenamefont {Chac{\'o}n}\ \emph {et~al.}()\citenamefont
  {Chac{\'o}n}, \citenamefont {Zhu}, \citenamefont {Kelly}, \citenamefont
  {Dauphin}, \citenamefont {Pisanty}, \citenamefont {Pic{\'o}n}, \citenamefont
  {Ticknor}, \citenamefont {Ciappina}, \citenamefont {Saxena},\ and\
  \citenamefont {Lewenstein}}]{chacon_observing_2018}%
  \BibitemOpen
  \bibfield  {author} {\bibinfo {author} {\bibfnamefont {A.}~\bibnamefont
  {Chac{\'o}n}}, \bibinfo {author} {\bibfnamefont {W.}~\bibnamefont {Zhu}},
  \bibinfo {author} {\bibfnamefont {S.~P.}\ \bibnamefont {Kelly}}, \bibinfo
  {author} {\bibfnamefont {A.}~\bibnamefont {Dauphin}}, \bibinfo {author}
  {\bibfnamefont {E.}~\bibnamefont {Pisanty}}, \bibinfo {author} {\bibfnamefont
  {A.}~\bibnamefont {Pic{\'o}n}}, \bibinfo {author} {\bibfnamefont
  {C.}~\bibnamefont {Ticknor}}, \bibinfo {author} {\bibfnamefont {M.~F.}\
  \bibnamefont {Ciappina}}, \bibinfo {author} {\bibfnamefont {A.}~\bibnamefont
  {Saxena}}, \ and\ \bibinfo {author} {\bibfnamefont {M.}~\bibnamefont
  {Lewenstein}},\ }\href {http://arxiv.org/abs/1807.01616v3} {\bibinfo
  {journal} {{arXiv}:1807.01616v3 [cond-mat, physics:quant-ph]}\ }\BibitemShut
  {NoStop}%
\bibitem [{\citenamefont {Dr\"ueke}\ and\ \citenamefont
  {Bauer}(2019)}]{DrueekeBauer2019}%
  \BibitemOpen
\bibfield  {journal} {  }\bibfield  {author} {\bibinfo {author} {\bibfnamefont
  {H.}~\bibnamefont {Dr\"ueke}}\ and\ \bibinfo {author} {\bibfnamefont
  {D.}~\bibnamefont {Bauer}},\ }\href {\doibase 10.1103/PhysRevA.99.053402}
  {\bibfield  {journal} {\bibinfo  {journal} {Phys. Rev. A}\ }\textbf {\bibinfo
  {volume} {99}},\ \bibinfo {pages} {053402} (\bibinfo {year}
  {2019})}\BibitemShut {NoStop}%
\bibitem [{\citenamefont {J\"ur\ss{}}\ and\ \citenamefont
  {Bauer}(2019)}]{JuerssBauer2019}%
  \BibitemOpen
  \bibfield  {author} {\bibinfo {author} {\bibfnamefont {C.}~\bibnamefont
  {J\"ur\ss{}}}\ and\ \bibinfo {author} {\bibfnamefont {D.}~\bibnamefont
  {Bauer}},\ }\href {\doibase 10.1103/PhysRevB.99.195428} {\bibfield  {journal}
  {\bibinfo  {journal} {Phys. Rev. B}\ }\textbf {\bibinfo {volume} {99}},\
  \bibinfo {pages} {195428} (\bibinfo {year} {2019})}\BibitemShut {NoStop}%
\bibitem [{\citenamefont {Luu}\ and\ \citenamefont
  {W{\"o}rner}(2018)}]{Luu2018}%
  \BibitemOpen
  \bibfield  {author} {\bibinfo {author} {\bibfnamefont {T.~T.}\ \bibnamefont
  {Luu}}\ and\ \bibinfo {author} {\bibfnamefont {H.~J.}\ \bibnamefont
  {W{\"o}rner}},\ }\href {\doibase 10.1038/s41467-018-03397-4} {\bibfield
  {journal} {\bibinfo  {journal} {Nature Communications}\ }\textbf {\bibinfo
  {volume} {9}},\ \bibinfo {pages} {916} (\bibinfo {year} {2018})}\BibitemShut
  {NoStop}%
\bibitem [{\citenamefont {Haldane}(1988)}]{Haldane_1988}%
  \BibitemOpen
  \bibfield  {author} {\bibinfo {author} {\bibfnamefont {F.~D.~M.}\
  \bibnamefont {Haldane}},\ }\href {\doibase 10.1103/PhysRevLett.61.2015}
  {\bibfield  {journal} {\bibinfo  {journal} {Phys. Rev. Lett.}\ }\textbf
  {\bibinfo {volume} {61}},\ \bibinfo {pages} {2015} (\bibinfo {year}
  {1988})}\BibitemShut {NoStop}%
\bibitem [{\citenamefont {Jotzu}\ \emph {et~al.}(2014)\citenamefont {Jotzu},
  \citenamefont {Messer}, \citenamefont {Desbuquois}, \citenamefont {Lebrat},
  \citenamefont {Uehlinger}, \citenamefont {Greif},\ and\ \citenamefont
  {Esslinger}}]{Jotzu2014}%
  \BibitemOpen
  \bibfield  {author} {\bibinfo {author} {\bibfnamefont {G.}~\bibnamefont
  {Jotzu}}, \bibinfo {author} {\bibfnamefont {M.}~\bibnamefont {Messer}},
  \bibinfo {author} {\bibfnamefont {R.}~\bibnamefont {Desbuquois}}, \bibinfo
  {author} {\bibfnamefont {M.}~\bibnamefont {Lebrat}}, \bibinfo {author}
  {\bibfnamefont {T.}~\bibnamefont {Uehlinger}}, \bibinfo {author}
  {\bibfnamefont {D.}~\bibnamefont {Greif}}, \ and\ \bibinfo {author}
  {\bibfnamefont {T.}~\bibnamefont {Esslinger}},\ }\href {\doibase
  10.1038/nature13915} {\bibfield  {journal} {\bibinfo  {journal} {Nature}\
  }\textbf {\bibinfo {volume} {515}},\ \bibinfo {pages} {237} (\bibinfo {year}
  {2014})}\BibitemShut {NoStop}%
\bibitem [{\citenamefont {Cao}\ \emph {et~al.}(2017)\citenamefont {Cao},
  \citenamefont {Zhao},\ and\ \citenamefont {Louie}}]{Cao2017}%
  \BibitemOpen
  \bibfield  {author} {\bibinfo {author} {\bibfnamefont {T.}~\bibnamefont
  {Cao}}, \bibinfo {author} {\bibfnamefont {F.}~\bibnamefont {Zhao}}, \ and\
  \bibinfo {author} {\bibfnamefont {S.~G.}\ \bibnamefont {Louie}},\ }\href
  {\doibase 10.1103/PhysRevLett.119.076401} {\bibfield  {journal} {\bibinfo
  {journal} {Phys. Rev. Lett.}\ }\textbf {\bibinfo {volume} {119}},\ \bibinfo
  {pages} {076401} (\bibinfo {year} {2017})}\BibitemShut {NoStop}%
\bibitem [{\citenamefont {Pantale{\'o}n}\ and\ \citenamefont
  {Xian}(2018)}]{PANTALEON2018191}%
  \BibitemOpen
  \bibfield  {author} {\bibinfo {author} {\bibfnamefont {P.~A.}\ \bibnamefont
  {Pantale{\'o}n}}\ and\ \bibinfo {author} {\bibfnamefont {Y.}~\bibnamefont
  {Xian}},\ }\href {\doibase https://doi.org/10.1016/j.physb.2017.11.040}
  {\bibfield  {journal} {\bibinfo  {journal} {Physica B: Condensed Matter}\
  }\textbf {\bibinfo {volume} {530}},\ \bibinfo {pages} {191 } (\bibinfo {year}
  {2018})}\BibitemShut {NoStop}%
\bibitem [{\citenamefont {Graf}\ and\ \citenamefont {Vogl}(1995)}]{Graf_1995}%
  \BibitemOpen
  \bibfield  {author} {\bibinfo {author} {\bibfnamefont {M.}~\bibnamefont
  {Graf}}\ and\ \bibinfo {author} {\bibfnamefont {P.}~\bibnamefont {Vogl}},\
  }\href {\doibase 10.1103/PhysRevB.51.4940} {\bibfield  {journal} {\bibinfo
  {journal} {Phys. Rev. B}\ }\textbf {\bibinfo {volume} {51}},\ \bibinfo
  {pages} {4940} (\bibinfo {year} {1995})}\BibitemShut {NoStop}%
\bibitem [{\citenamefont {Kuzemsky}(2011)}]{Review_Transport}%
  \BibitemOpen
  \bibfield  {author} {\bibinfo {author} {\bibfnamefont {A.~L.}\ \bibnamefont
  {Kuzemsky}},\ }\href {\doibase 10.1142/S0217979211059012} {\bibfield
  {journal} {\bibinfo  {journal} {International Journal of Modern Physics B}\
  }\textbf {\bibinfo {volume} {25}},\ \bibinfo {pages} {3071} (\bibinfo {year}
  {2011})}\BibitemShut {NoStop}%
\bibitem [{\citenamefont {Bandrauk}\ \emph {et~al.}(2009)\citenamefont
  {Bandrauk}, \citenamefont {Chelkowski}, \citenamefont {Diestler},
  \citenamefont {Manz},\ and\ \citenamefont {Yuan}}]{Bandrauk2009}%
  \BibitemOpen
  \bibfield  {author} {\bibinfo {author} {\bibfnamefont {A.~D.}\ \bibnamefont
  {Bandrauk}}, \bibinfo {author} {\bibfnamefont {S.}~\bibnamefont
  {Chelkowski}}, \bibinfo {author} {\bibfnamefont {D.~J.}\ \bibnamefont
  {Diestler}}, \bibinfo {author} {\bibfnamefont {J.}~\bibnamefont {Manz}}, \
  and\ \bibinfo {author} {\bibfnamefont {K.-J.}\ \bibnamefont {Yuan}},\ }\href
  {\doibase 10.1103/PhysRevA.79.023403} {\bibfield  {journal} {\bibinfo
  {journal} {Phys. Rev. A}\ }\textbf {\bibinfo {volume} {79}},\ \bibinfo
  {pages} {023403} (\bibinfo {year} {2009})}\BibitemShut {NoStop}%
\bibitem [{\citenamefont {Baggesen}\ and\ \citenamefont
  {Madsen}(2011)}]{Baggesen_2011}%
  \BibitemOpen
  \bibfield  {author} {\bibinfo {author} {\bibfnamefont {J.~C.}\ \bibnamefont
  {Baggesen}}\ and\ \bibinfo {author} {\bibfnamefont {L.~B.}\ \bibnamefont
  {Madsen}},\ }\href {\doibase 10.1088/0953-4075/44/11/115601} {\bibfield
  {journal} {\bibinfo  {journal} {Journal of Physics B: Atomic, Molecular and
  Optical Physics}\ }\textbf {\bibinfo {volume} {44}},\ \bibinfo {pages}
  {115601} (\bibinfo {year} {2011})}\BibitemShut {NoStop}%
\bibitem [{\citenamefont {Bauer}\ \emph {et~al.}()\citenamefont {Bauer},
  \citenamefont {Bauke}, \citenamefont {Brabec}, \citenamefont {Fennel},
  \citenamefont {{McDonald}}, \citenamefont {Milošević}, \citenamefont
  {Pabst}, \citenamefont {Peltz}, \citenamefont {Pöplau}, \citenamefont
  {Santra},\ and\ \citenamefont {Varin}}]{bauer_computational_2017}%
  \BibitemOpen
  \bibfield  {author} {\bibinfo {author} {\bibfnamefont {D.}~\bibnamefont
  {Bauer}}, \bibinfo {author} {\bibfnamefont {H.}~\bibnamefont {Bauke}},
  \bibinfo {author} {\bibfnamefont {T.}~\bibnamefont {Brabec}}, \bibinfo
  {author} {\bibfnamefont {T.}~\bibnamefont {Fennel}}, \bibinfo {author}
  {\bibfnamefont {C.~R.}\ \bibnamefont {{McDonald}}}, \bibinfo {author}
  {\bibfnamefont {D.~B.}\ \bibnamefont {Milošević}}, \bibinfo {author}
  {\bibfnamefont {S.}~\bibnamefont {Pabst}}, \bibinfo {author} {\bibfnamefont
  {C.}~\bibnamefont {Peltz}}, \bibinfo {author} {\bibfnamefont
  {G.}~\bibnamefont {Pöplau}}, \bibinfo {author} {\bibfnamefont
  {R.}~\bibnamefont {Santra}}, \ and\ \bibinfo {author} {\bibfnamefont
  {C.}~\bibnamefont {Varin}},\ }\href
  {https://books.google.de/books?id=kwXEDgAAQBAJ} {\emph {\bibinfo {title}
  {Computational Strong-Field Quantum Dynamics: Intense Light-Matter
  Interactions}}},\ edited by\ \bibinfo {editor} {\bibfnamefont
  {D.}~\bibnamefont {Bauer}},\ De Gruyter Textbook\ (\bibinfo  {publisher} {De
  Gruyter})\BibitemShut {NoStop}%
\bibitem [{\citenamefont {Cooper}\ \emph {et~al.}(2012)\citenamefont {Cooper},
  \citenamefont {D'Anjou}, \citenamefont {Ghattamaneni}, \citenamefont
  {Harack}, \citenamefont {Hilke}, \citenamefont {Horth}, \citenamefont
  {Majlis}, \citenamefont {Massicotte}, \citenamefont {Vandsburger},
  \citenamefont {Whiteway},\ and\ \citenamefont {Yu}}]{Cooper_2012}%
  \BibitemOpen
  \bibfield  {author} {\bibinfo {author} {\bibfnamefont {D.~R.}\ \bibnamefont
  {Cooper}}, \bibinfo {author} {\bibfnamefont {B.}~\bibnamefont {D'Anjou}},
  \bibinfo {author} {\bibfnamefont {N.}~\bibnamefont {Ghattamaneni}}, \bibinfo
  {author} {\bibfnamefont {B.}~\bibnamefont {Harack}}, \bibinfo {author}
  {\bibfnamefont {M.}~\bibnamefont {Hilke}}, \bibinfo {author} {\bibfnamefont
  {A.}~\bibnamefont {Horth}}, \bibinfo {author} {\bibfnamefont
  {N.}~\bibnamefont {Majlis}}, \bibinfo {author} {\bibfnamefont
  {M.}~\bibnamefont {Massicotte}}, \bibinfo {author} {\bibfnamefont
  {L.}~\bibnamefont {Vandsburger}}, \bibinfo {author} {\bibfnamefont
  {E.}~\bibnamefont {Whiteway}}, \ and\ \bibinfo {author} {\bibfnamefont
  {V.}~\bibnamefont {Yu}},\ }\href@noop {} {\bibfield  {journal} {\bibinfo
  {journal} {ISRN Condensed Matter Physics}\ }\textbf {\bibinfo {volume}
  {2012}},\ \bibinfo {pages} {501686} (\bibinfo {year} {2012})}\BibitemShut
  {NoStop}%
\bibitem [{\citenamefont {Hansen}\ \emph {et~al.}(2017)\citenamefont {Hansen},
  \citenamefont {Deffge},\ and\ \citenamefont {Bauer}}]{PhysRevA.96.053418}%
  \BibitemOpen
  \bibfield  {author} {\bibinfo {author} {\bibfnamefont {K.~K.}\ \bibnamefont
  {Hansen}}, \bibinfo {author} {\bibfnamefont {T.}~\bibnamefont {Deffge}}, \
  and\ \bibinfo {author} {\bibfnamefont {D.}~\bibnamefont {Bauer}},\ }\href
  {\doibase 10.1103/PhysRevA.96.053418} {\bibfield  {journal} {\bibinfo
  {journal} {Phys. Rev. A}\ }\textbf {\bibinfo {volume} {96}},\ \bibinfo
  {pages} {053418} (\bibinfo {year} {2017})}\BibitemShut {NoStop}%
\end{thebibliography}%

\end{document}